\documentclass[sn-mathphys-num]{sn-jnl}

\usepackage{graphicx}%
\usepackage{multirow}%
\usepackage{amsmath,amssymb,amsfonts}%
\usepackage{amsthm}%
\usepackage{mathrsfs}%
\usepackage[title]{appendix}%
\usepackage{xcolor}%
\usepackage{textcomp}%
\usepackage{manyfoot}%
\usepackage{booktabs}%
\usepackage{algorithm}%
\usepackage{algorithmicx}%
\usepackage{algpseudocode}%
\usepackage{listings}%

\theoremstyle{thmstyleone}%
\theoremstyle{thmstyletwo}%

\theoremstyle{thmstylethree}%

\begin{document}
	
	\title[\textbf{A system of }$\mathbf{2}$\textbf{ nonlinearly coupled ODEs }				\textbf{which is explicitly solvable and possibly isochronous}		\textbf{provided its coefficients are suitably restricted}]{\textbf{A system of }$\mathbf{2}$\textbf{ nonlinearly coupled ODEs }				\textbf{which is explicitly solvable and possibly isochronous}		\textbf{provided its coefficients are suitably restricted}}

	\author*[1,2,3]{\fnm{Fabio} \sur{Briscese}}\email{fabio.briscese@uniroma3.it}\equalcont{These authors contributed equally to this work.}

	\author[4,5]{\fnm{Francesco} \sur{Calogero}}\email{francesco.calogero@uniroma1.it, francesco.calogero@infn.roma1.it}
	\equalcont{These authors contributed equally to this work.}
	
	\author[6]{\fnm{Farrin} \sur{Payandeh}}\email{f\_payandeh@pnu.ac.ir, farrinpayandeh@yahoo.com.}
	\equalcont{These authors contributed equally to this work.}
	
	\affil*[1]{\orgdiv{Dipartimento di Architettura}, \orgname{Universit\`{a} Roma Tre}, \orgaddress{\street{Via Aldo Manuzio, 68L}, \city{Rome}, \postcode{00153}, \country{Italy}}}
	
	\affil[2]{\orgdiv{Gruppo
			Nazionale di Fisica Matematica}, \orgname{Istituto Nazionale di Alta Matematica Francesco Severi}, \orgaddress{\street{P. Aldo Moro 5}, \city{Rome}, \postcode{00185}, \country{Italy}}}
	
	\affil[3]{\orgname{Istituto Nazionale di Fisica Nucleare}, \orgdiv{Sezione di Roma 3}, \orgaddress{\street{ Via della Vasca Navale, 84}, \city{Rome}, \postcode{00146},  \country{Italy}}}

	\affil[4]{\orgdiv{Dipartimento di Fisica}, \orgname{Sapienza Università di Roma}, \orgaddress{\street{ P. Aldo Moro 5}, \city{Rome}, \postcode{00185},  \country{Italy}}}
	
	\affil[5]{\orgname{Istituto Nazionale di Fisica Nucleare}, \orgdiv{Sezione di Roma 1}, \orgaddress{\street{ P. Aldo Moro, 5}, \city{Rome}, \postcode{00185},  \country{Italy}}}

	\affil[6]{\orgdiv{Department of Physics}, \orgname{Payame Noor University (PNU)},  \orgaddress{\street{ PO\ BOX 19395-3697}, \city{Tehran},  \country{Iran}}}

	\abstract{In this paper we discuss some remarkable properties of the \textit{autonomous} system of 2 \textit{first-order} Ordinary Differential Equations (ODEs), which equates the derivatives $\dot{x}_n(t)$ ($n = 1, 2$) of the 2 dependent variables $x_n(t)$ to the \textit{ratios} of \textit{polynomials} (with constant coefficients) in the 2 variables $x_n (t)$: each of the 2 (\textit{a priori} different) \textit{polynomials} $P_3^{(n)}(x_1, x_2)$ in the 2 numerators is of degree 3; the 2 denominators are instead given by the same polynomial $P_1(x_1, x_2)$ of degree 1. Hence this system features 23 \textit{a priori} \textit{arbitrary} input numbers, namely the 23 \textit{coefficients} defining these 3 polynomials. Our main finding is to show that if these 23 \textit{coefficients} are given by 23 (\textit{explicitly provided}) formulas in terms of 15 \textit{a priori} \textit{arbitrary parameters}, then the \textit{initial values} problem (with \textit{arbitrary} initial data $x_n (0)$) for this dynamical system can be \textit{explicitly} solved. We also show that it is possible (with the help of \textbf{Mathematica}) to identify 12 \textit{explicit} \textit{constraints} on these 23 \textit{coefficients}, which are \textit{sufficient} to guarantee that this system belongs to the class of systems we are focusing on. Several such \textit{explicitly} solvable systems of ODEs are treated (including the subcase with $P_1(x_1, x_2) = 1$, implying that the right-hand sides of the ODEs are \textit{just cubic polynomials}: no denominators!). Examples of the solutions of several of these systems are reported and displayed, including cases in which the solutions are \textit{isochronous}.}

	
	
	
	\maketitle

\section{Introduction}

In this paper we focus on the following system of $2$ first-order
\textit{nonlinearly-coupled} Ordinary Differential Equations (ODEs):
\begin{subequations}
\label{ODEs}%
\begin{equation}
\dot{x}_{n}=P_{3}^{\left(  n\right)  }\left(  x_{1},x_{2}\right)
/P_{1}\left(  x_{1},x_{2}\right)  ~,~~~n=1,2~, \label{xx12dot}%
\end{equation}
where $P_{3}^{\left(  n\right)  }\left(  x_{1},x_{2}\right)  $ (with $n=1,2$)
are the following $2$ \textit{polynomials} of \textit{third }degree:
\begin{align}
&  P_{3}^{\left(  1\right)  }\left(  x_{1},x_{2}\right)
=f_{9}\left(  x_{1}\right)  ^{3}+f_{8}\left(  x_{1}\right)  ^{2}x_{2}%
+f_{7}x_{1}\left(  x_{2}\right)  ^{2}+f_{6}\left(  x_{2}\right)
^{3}\nonumber\\
&  +f_{5}\left(  x_{1}\right)  ^{2}+f_{4}x_{1}x_{2}+f_{3}\left(  x_{2}\right)
^{2}+f_{2}x_{1}+f_{1}x_{2}+f_{0}~,
\end{align}%
\begin{align}
&  P_{3}^{\left(  2\right)  }\left(  x_{1},x_{2}\right)
=g_{9}\left(  x_{1}\right)  ^{3}+g_{8}\left(  x_{1}\right)  ^{2}x_{2}%
+g_{7}x_{1}\left(  x_{2}\right)  ^{2}+g_{6}\left(  x_{2}\right)
^{3}\nonumber\\
&  +g_{5}\left(  x_{1}\right)  ^{2}+g_{4}x_{1}x_{2}+g_{3}\left(  x_{2}\right)
^{2}+g_{2}x_{1}+g_{1}x_{2}+g_{0}~,
\end{align}
and $P_{1}\left(  x_{1},x_{2}\right)  $ is the following \textit{polynomial}
of \textit{first }degree:
\begin{equation}
P_{1}\left(  x_{1},x_{2}\right)  =h_{1}x_{1}+h_{2}x_{2}+h_{0}~.
\end{equation}

\textbf{Notation}. Hereafter the indices $n$ and $m$ take the $2$ values
$1,2$, the index $k$ takes the $10$ integer values from $9$ to $0$, and the
index $\ell$ takes the $3$ values $2,1,0$; the symbol $t$ denotes the
\textit{independent variable}; the quantities $x_{n}\left(  t\right)  $ and
(see below) $y_{n}\left(  t\right)  $ are \textit{dependent variables }(but
often their $t$-dependence will \textit{not} be \textit{explicitly} displayed,
see for instance above); and a superimposed dot denotes the $t$-derivative of
these quantities, $\dot{x}_{n}\left(  t\right)  \equiv dx_{n}\left(  t\right)
/dt,~\dot{y}_{n}\left(  t\right)  \equiv dy_{n}\left(  t\right)  /dt$. We
will\textit{ }occasionally assume that the independent variable $t$ is "time"
(but this is of course only a matter of terminology). The $2\cdot10+3=23$
\textit{coefficients} $f_{k}$, $g_{k}$ and $h_{\ell}$ are $t$-independent quantities,
and we consider them to be \textit{generically} assigned (\textit{unless
otherwise indicated}); their number might of course be trivially reduced from
$23$ to $22$ by dividing \textit{all} of them by \textit{anyone} of them,
which is thereby reduced to \textit{unity} (unless it vanished to begin with),
since this operation clearly leaves invariant the system of ODEs (\ref{ODEs}).
In this paper we \textit{generally} focus on the case in which the
\textit{variables} and the \textit{coefficients} are \textit{all real
}numbers, because this is the case of main \textit{applicative} interest; but
of course in the context of some computations \textit{complex} numbers may
also play a role; and we use the symbol $\mathbf{i}$ to denote the
\textit{imaginary unit }(so $\mathbf{i}^{2}=-1$). The $15=3+2+1+2+1$ $+4+2$
\textit{parameters} $\alpha_{\ell}$, $\beta_{n}$, $\gamma$, $\lambda_{n}$, $\mu$,
$\eta_{nm},$ $\xi_{n}$ used below are also all $t$-independent. $\blacksquare$

Systems of nonlinear ODEs such as that considered in this paper have been
studied for centuries by mathematicians; the relevant literature includes
\textit{thousands} of scientific papers; see, for instance, the classic paper
\cite{G1960} and the books \cite{I1956}, \cite{S1994}, \cite{DLA2006},
\cite{FC2012}, \cite{PZ2017}, as well as the website \textbf{EqWorld} (and the
papers and books referred to there).

These systems of nonlinear ODEs play a key role in many \textit{applicative}
contexts; this is also the case for the specific system treated in this paper
(for instance in simple models of population dynamics, epidemiology,
economics, engineering,...); but below we focus on the \textit{mathematical}
analysis of this system of ODEs; with some emphasis on certain mathematical
developments which might be important for its \textit{applicative} usefulness. We  also emphasize that, with this approach, we are able to identify subclasses of the system of ODEs (\ref{ODEs}) which feature 	\textit{periodic}, in fact \textit{isochronous}, solutions.

\textbf{Remark 1-1}. Since in the title of this paper we used the terminology ``system of $2$ nonlinearly coupled ODEs which is \textit{explicitly
solvable}", it is perhaps appropriate that we recall what is usually meant by
it: such systems are those whose \textit{initial values problem} (for
\textit{arbitrarily assigned initial data}) can be \textit{explicitly solved}
in terms of an \textit{explicit} combination of \textit{elementary functions} of the \textit{independent variable} $t$,
involving a number of (of course \textit{constant}) \textit{parameters} which can
themselves be \textit{computed}, in terms of the (of course \textit{constant})
\textit{coefficients} defining the system of ODEs and of the \textit{initial
data}, via purely \textit{algebraic operations} (typically eventually
reducible to finding the roots of given polynomials; hence not themselves
necessarily \textit{explicitly performable}; although, in the specific examples
considered below, this shall indeed be the case). Let us also remind the
reader of a well-known notion. Such a definition is \textit{more stringent}
than the definition of \textit{integrable systems}, which implies the
\textit{existence} of the \textit{maximal} \textit{number }of
\textit{constants of motion} (which, for systems such as that considered in
this paper, is in fact just $1$). So, \textit{explicit solvability} implies
\textit{integrability }(as usually defined in terms of the \textit{number}
of\textit{ constants of motion}, or, equivalently, of \textit{conserved
quantities}); while the \textit{reverse} is \textit{not true}. And it is
certainly more useful in terms of \textit{applicability}, when finally the
issue is reduced to the comparison of \textit{experimental} data (obtained or
to be obtained) with \textit{numerical} computations of findings produced by
\textit{theoretical} models. $\blacksquare$

\bigskip

\section{Main results}\label{Section main results}

This \textbf{Section 2} is divided in $4$ \textbf{Subsections}. In
\textbf{Subsection 2.1} we introduce and discuss a \textit{simple} system of
$2$ \textit{linear first-order} ODEs, which is of course \textit{explicitly
solvable}. In \textbf{Subsection 2.2} we introduce a simple,
\textit{explicitly invertible} relation among the $2$ dependent variables
$x_{n}\left(  t\right)  $ of our dynamical system (\ref{ODEs}) and the $2$
dependent variables $y_{n}\left(  t\right)  $ of the \textit{explicitly}
\textit{solvable} system described in \textbf{Subsection 2.1}; and we show that this
simple relation among the variables $x_{n}\left(  t\right)  $ and
$y_{n}\left(  t\right)  $ entails that, if the variables $y_{n}\left(
t\right)  $ evolve according to the simple \textit{solvable} system detailed in
\textbf{Subsection 2.1}, the $2$ variables $x_{n}\left(  t\right)  $ indeed
evolve according to our system of ODEs (\ref{ODEs}) (with suitable
restrictions on the \textit{coefficients}, as detailed below). In \textbf{Subsection
2.3} we treat the interesting subcase of the system of ODEs (\ref{ODEs})
which\textbf{---}provided their solutions do satisfy, for \textit{all positive} times $t$,
the \textit{initial values} problem (with \textit{arbitrary real initial
data}) of the system of ODEs (\ref{ODEs})---are then \textit{isochronous} (i.
e., they are \textit{all} periodic with the \textit{same} period $T$
\textit{independent} of the initial data). In \textbf{Subsection 2.4 }we
report the \textit{constraints }on the \textit{coefficients} of the system of ODEs (\ref{ODEs}) that correspond to the explicitly solvable case studied in this paper, and in \textbf{Subsection 2.5 }we report additional properties of such systems. In \textbf{Section 3} we report and display 5 \textit{explicit} examples. Finally, in \textbf{Section 4} we tersely present some concluding remarks.

\bigskip

\subsection{An auxiliary, solvable system of $2$ linear ODEs}\label{An auxiliary, solvable system of $2$ linear ODEs}

In this \textbf{Subsection 2.1 }we introduce a simple system of $2$
\textit{linear} ODEs and report the \textit{explicit} solution of its
\textit{initial values} problem. These are of course \textit{elementary} results.

Let us then assume that the $2$ variables $y_{n}\left(  t\right)  $ evolve
(linearly!) as follows:
\end{subequations}
\begin{equation}
\dot{y}_{1}=\eta_{11}y_{1}+\eta_{12}y_{2}+\xi_{1}~,~~~\dot{y}_{2}=\eta
_{21}y_{1}+\eta_{22}y_{2}+\xi_{2}~. \label{ydot}%
\end{equation}
This system involves $4+2=6$ \textit{a priori arbitrary} \textit{parameters} $\eta
_{nm}$ and $\xi_{n}$. The solution of the initial values problem of this
system of ODEs (\ref{ydot}) reads as follows (provided $\tilde{\eta}\neq 0$, see below eq. (\ref{eta tilde})):
\begin{subequations}
\label{Solyn(t)}%
\begin{equation}
y_{n}\left(  t\right)  =v_{n}^{\left(  +\right)  }\exp\left[  \phi^{\left(
+\right)  }t\right]  +v_{n}^{\left(  -\right)  }\exp\left[  \phi^{\left(
-\right)  }t\right]  +u_{n}~;
\end{equation}%
\begin{equation}\label{u1}
u_{1}=\left(  \eta_{12}\xi_{2}-\eta_{22}\xi_{1}\right)
/\eta~,
\end{equation}%
\begin{equation}\label{u2}
u_{2}=\left(  \eta_{21}\xi_{1}-\eta_{11}\xi_{2}\right)
/\eta~,
\end{equation}%
\begin{equation}
\eta=\eta_{11}\eta_{22}-\eta_{12}\eta_{21}~;
\end{equation}
~%
\begin{equation}\label{phi pm}
\phi^{\left(  \pm\right)  }=\bar{\eta}\pm\tilde{\eta}~,
\end{equation}%
\begin{equation}\label{eta bar}
\bar{\eta}\equiv\left(  \eta_{11}+\eta_{22}\right)  /2~,
\end{equation}

\begin{equation}	\label{eta tilde}
\tilde{\eta}=\sqrt{\left(  \eta_{11}-\eta_{22}\right)
^{2}+4\eta_{12}\eta_{21}}~/2;
\end{equation}
\begin{equation}
v_{1}^{\left(  +\right)  }=\left\{  \left(  -\eta_{11}%
+\eta_{22}-2\tilde{\eta}\right)  \left[  u_{1}-y_{1}\left(  0\right)  \right]
-2\eta_{12}\left[  u_{2}-y_{2}\left(  0\right)  \right]  \right\}  /\left(
4\tilde{\eta}\right)  ~,
\end{equation}%
\begin{equation}
v_{1}^{\left(  -\right)  }=\left\{  \left(  \eta_{11}%
-\eta_{22}-2\tilde{\eta}\right)  \left[  u_{1}-y_{1}\left(  0\right)  \right]
+2\eta_{12}\left[  u_{2}-y_{2}\left(  0\right)  \right]  \right\}  /\left(
4\tilde{\eta}\right)  ~,
\end{equation}%
\begin{equation}
v_{2}^{\left(  +\right)  }=\left\{  \left(  \eta_{11}%
-\eta_{22}-2\tilde{\eta}\right)  \left[  u_{2}-y_{2}\left(  0\right)  \right]
-2\eta_{21}\left[  u_{1}-y_{1}\left(  0\right)  \right]  \right\}  /\left(
4\tilde{\eta}\right)  ~,
\end{equation}%
\begin{equation}
v_{2}^{\left(  -\right)  }=\left\{  \left(  -\eta_{11}%
+\eta_{22}-2\tilde{\eta}\right)  \left[  u_{2}-y_{2}\left(  0\right)  \right]
+2\eta_{21}\left[  u_{1}-y_{1}\left(  0\right)  \right]  \right\}  /\left(
4\tilde{\eta}\right)  ~.
\end{equation}

It is moreover clear that, if the $4$ coefficients $\eta_{nm}$ satisfy the
following $2$ simple \textit{restrictions} (an \textit{equality} and an
\textit{inequality})
\end{subequations}
\begin{equation}\label{restriction eta}
\eta_{22}=-\eta_{11}~,~~~\left(  \eta_{11}\right)  ^{2}+\eta_{12}\eta
_{21}=-\omega^{2}<0~,
\end{equation}
where  the quantity $\omega$ is a \textit{real} number, which for definiteness is hereafter assumed to be
\textit{positive}, then this system of ODEs is \textit{isochronous}, namely \textit{all} its
solutions are clearly \textit{periodic} with the \textit{same} period:
\begin{subequations}\label{isochronous y solutions}
\begin{equation}
\phi^{\left(  +\right)  }=\mathbf{i}\omega~,~~~\phi^{\left(  -\right)
}=-\mathbf{i}\omega~,
\end{equation}%
\begin{equation}\label{omega}
\omega=\sqrt{-\left[  \left(  \eta_{11}\right)  ^{2}+\eta_{12}\eta
_{21}\right]  }>0~,
\end{equation}%
\begin{equation}\label{definition period}
y_{n}\left(  t+T\right)  =y_{n}\left(  t\right)  ~,~~~T=2\pi/\omega~.
\end{equation}%
\begin{align}
&  y_{1}\left(  t\right)  =u_{1}+\left[  y_{1}\left(  0\right)  -u_{1}\right]
\cos\left(  \omega t\right)  +~\nonumber\\
&  +\left\{  \left[  \eta_{11}\left[  y_{1}\left(  0\right)  -u_{1}\right]
+\eta_{12}\left[  y_{2}\left(  0\right)  -u_{2}\right]  \right]  \right\}
\sin\left(  \omega t\right)  /\omega~,
\end{align}%
\begin{align}
&  y_{2}\left(  t\right)  =u_{2}+\left[  y_{2}\left(  0\right)  -u_{2}\right]
\cos\left(  \omega t\right) \nonumber\\
&  +\left\{  \left[  \eta_{22}\left[  y_{2}\left(  0\right)  -u_{2}\right]
+\eta_{21}\left[  y_{1}\left(  0\right)  -u_{1}\right]  \right]  \right\}
\sin\left(  \omega t\right)  /\omega~.
\end{align}
This special case is further discussed in \textbf{Subsection 2.3}.

\bigskip

\subsection{A simple relation among the pairs of variables $x_{n}\left(
t\right)  $ and $y_{n}\left(  t\right)  $}\label{A simple relation among the pair of variables}

In this\textbf{\ Subsection 2.2} we relate the $2$ \textit{new} dependent
variables $y_{n}\left(  t\right)  $ (introduced in \textbf{Subsection
2.1}),\textbf{\ }to the\textbf{\ }$2$ dependent variables $x_{n}\left(
t\right)  $ (introduced in \textbf{Section 1}), by setting
\end{subequations}
\begin{subequations}
\label{xyt}%
\begin{equation}
y_{1}=\alpha_{0}\left(  x_{1}\right)  ^{2}+\alpha_{1}x_{1}x_{2}+\alpha
_{2}\left(  x_{2}\right)  ^{2}+\beta_{1}x_{1}+\beta_{2}x_{2}+\gamma~,
\end{equation}%
\begin{equation}
y_{2}=\lambda_{1}x_{1}+\lambda_{2}x_{2}+\mu~.
\end{equation}
This \textit{ansatz} involves the $3+2+1+2+1=9$ \textit{a priori arbitrary}
parameters $\alpha_{\ell}$, $\beta_{n}$, $\gamma$, $\lambda_{n}$, $\mu$ (hereafter we consider this 9 parameters to take generic values. Special cases in which some of them take special values---for instance, vanish---are discussed below).

It is easily seen that this \textit{ansatz} can be explicitly inverted:
\end{subequations}
\begin{subequations}
\label{Solx1x2(t)}%
\begin{equation}
x_{1}\left(  t\right)  =\left[  -C_{1}\left(  t\right)  \pm S\left(  t\right)
\right]  /\left(  2C_{2}\right)  \equiv X_{1}^{\left(  \pm\right)  }\left(
y_{1}(t),y_{2}(t)\right)  ~, \label{x1(t)}%
\end{equation}%
\begin{equation}
x_{2}\left(  t\right)  =\left[  y_{2}\left(  t\right)
-\lambda_{1}X_{1}^{\left(  \pm\right)  }\left[  y_{1}(t),y_{2}(t)\right]
-\mu\right]  /\lambda_{2}\equiv X_{2}^{\left(  \pm\right)  }\left(
y_{1}(t),y_{2}(t)\right)  ~; \label{x2(t)}%
\end{equation}%
\begin{equation}
S\left(  t\right)  =\sqrt{\left[  C_{1}\left(  t\right)
\right]  ^{2}-4C_{2}C_{0}\left(  t\right)  }~; \label{S(t)}%
\end{equation}%
\begin{equation}\label{C2}
C_{2}=\alpha_{0}\left(  \lambda_{2}\right)  ^{2}-\alpha
_{1}\lambda_{1}\lambda_{2}+\alpha_{2}\left(  \lambda_{1}\right)^{2}~,
\end{equation}%
\begin{equation}
C_{1}\left(  t\right)  =B_{0}+B_{1}y_{2}\left(  t\right)  ~,
\end{equation}%
\begin{equation}
C_{0}\left(  t\right)  =A_{0}+A_{1}y_{1}\left(  t\right)
+A_{2}y_{2}\left(  t\right)  +\alpha_{2}\left[  y_{2}\left(  t\right)
\right]  ^{2}~;
\end{equation}%
\begin{equation}
B_{0}=-\alpha_{1}\lambda_{2}\mu+2\alpha_{2}\lambda_{1}%
\mu+\beta_{1}\lambda_{2}^{2}-\beta_{2}\lambda_{1}\lambda_{2},
\end{equation}%
\begin{equation}
B_{1}=\alpha_{1}\lambda_{2}-2\alpha_{2}\lambda_{1}~;
\end{equation}%
\begin{equation}
A_{0}=\alpha_{2}\mu^{2}-\beta_{2}\lambda_{2}\mu
+\gamma\left(  \lambda_{2}\right)  ^{2}~,
\end{equation}%
\begin{equation}
A_{1}=-\left(  \lambda_{2}\right)  ^{2}~,
\end{equation}%
\begin{equation}
A_{2}=\beta_{2}\lambda_{2}-2\alpha_{2}\mu~,
\end{equation}
\end{subequations}
provided that  $C_2 \neq 0$; otherwise

\begin{subequations}
	\label{Solx1x2(t) 2}%
	\begin{equation}
		x_{1}\left(  t\right)  =  -C_{0}\left(  t\right)/C_{1}\left(  t\right)   ~, \label{x1(t) 2}%
	\end{equation}%
	\begin{equation}
		x_{2}\left(  t\right)  =\left[  y_{2}\left(  t\right)
		-\lambda_{1}x_{1}\left(  t\right)
		-\mu\right]  /\lambda_{2} \, .
	\end{equation}
\end{subequations}

\textbf{Remark 2.2-1.} One might wonder why there \textit{seems} to be some
\textit{ambiguity} in the transition from the variables $y_{n}\left(
t\right)  $ to the variables $x_{n}\left(  t\right)  ,$ see the $\pm$ signs in
(\ref{x1(t)}). But this \textit{ambiguity} is \textit{removed} at $t=0$ by
using the \textit{unambiguous} relations (\ref{xyt}) to choose the right sign; and then, by
\textit{continuity}, keeping the same sign as $t$ increases. Of course, due to the
\textit{nonlinear} character of the system of ODEs (\ref{ODEs}), $2$ phenomena
may interrupt the \textit{unique} determination of the solution of its
\textit{initial values} problem. One such possibility is, that at some
(positive, finite) value $t=t_{\infty}$, a \textit{blow-up} of the solutions
occur; but this does not seem to be the case here, see (\ref{Solx1x2(t)}) and
the preceding \textbf{Subsection 2.1}. Another possibility is, that at some
value $t=t_{S}$ the square-root $S\left(  t\right)  $ (see (\ref{S(t)}))
vanish (so that $S\left(  t_{S}\right)  =0$); then the solutions
\textit{loose} the property to be thereafter \textit{uniquely} determined by
the system of ODEs (\ref{ODEs}); this is due to the fact that, at $t=t_{S}$,
the \textit{denominator} in the right-hand side\ of the ODEs (\ref{ODEs})
vanishes, so that the system of ODEs (\ref{ODEs}) encounters at $t=t_{S}$ a
\textit{singularity} which impedes the \textit{unique} determination of the
$t$-evolution of the dependent variables $x_{n}\left(  t\right)  $
\textit{beyond }that value of the independent variable $t$. $\blacksquare$

\textbf{Remark 2.2-2}. Let us emphasize that the \textit{first} (no blow-up) of the $2$ properties
mentioned in the preceding \textbf{Remark 2.2-1 }could by no means be guessed
by looking at the right-hand side of that system; indeed it is obviously
\textit{not} a property of \textit{all} the solutions of that system for any
\textit{arbitrary} assignment of its $23$ coefficients: it is a
\textit{special} property of the subclass of that system which we identify in
this paper (perhaps making that subclass quite interesting in applicative
contexts; although some solutions of those system may still cease to satisfy
the system of ODEs (\ref{ODEs}) at some \textit{finite} value of the independent
variable due to the \textit{second} (loss of determination) phenomenon mentioned in \textbf{Remark 2.2-1}).
$\blacksquare$

It is moreover a matter of trivial if tedious algebra (facilitated by using
appropriately \textbf{Mathematica}) to check that, if the variables
$y_{n}\left(  t\right)  $ evolve in $t$ according to the system of ODEs
(\ref{ydot}) and are related to the variables $x_{n}\left(  t\right)  $ by the
$2$ relations (\ref{xyt}), the variables $x_{n}\left(  t\right)  $ do indeed
satisfy our system of ODEs (\ref{ODEs}), provided the $23$ \textit{coefficients}
$f_{k}$, $g_{k}$ and $h_{\ell}$ are given in terms of the $4+2+3+2+1+2+1=15$
\textit{parameters} $\eta_{nm},$ $\xi_{n},$ $\alpha_{\ell},$ $\beta_{n},$ $\gamma,$
$\lambda_{n},$ $\mu$ by the following $23$ \textit{explicit} formulas:
\begin{subequations}
\label{f1f2g}%
\begin{align}\label{f0}
f_{0}  &  =-\beta_{2}\gamma\eta_{21}-\beta_{2}\eta_{22}\mu-\beta_{2}\xi
_{2}+\gamma\eta_{11}\lambda_{2}\nonumber\\
&  +\eta_{12}\lambda_{2}\mu+\lambda_{2}\xi_{1}~,
\end{align}%
\begin{align}\label{f1}
f_{1}  &  =-2\alpha_{2}\gamma\eta_{21}-2\alpha_{2}\eta_{22}%
\mu-2\alpha_{2}\xi_{2}+\beta_{2}\eta_{11}\lambda_{2}\nonumber\\
&  -\beta_{2}\eta_{22}\lambda_{2}-\left(  \beta_{2}\right)  ^{2}\eta_{21}%
+\eta_{12}\left(  \lambda_{2}\right)  ^{2}~,
\end{align}%
\begin{align}\label{f2}
f_{2}  &  =-\alpha_{1}\gamma\eta_{21}-\alpha_{1}\eta_{22}%
\mu-\alpha_{1}\xi_{2}-\beta_{2}\eta_{22}\lambda_{1}\nonumber\\
&  +\beta_{1}\eta_{11}\lambda_{2}-\beta_{1}\beta_{2}\eta_{21}+\eta_{12}%
\lambda_{1}\lambda_{2}~,
\end{align}%
\begin{equation}\label{f3}
f_{3}=-3\alpha_{2}\beta_{2}\eta_{21}+\alpha_{2}\eta_{11}%
\lambda_{2}-2\alpha_{2}\eta_{22}\lambda_{2}~,
\end{equation}%
\begin{align}\label{f4}
f_{4}  &  =-2\alpha_{2}\beta_{1}\eta_{21}-2\alpha_{1}\beta
_{2}\eta_{21}-2\alpha_{2}\eta_{22}\lambda_{1}\nonumber\\
&  +\alpha_{1}\eta_{11}\lambda_{2}-\alpha_{1}\eta_{22}\lambda_{2}~,
\end{align}%
\begin{equation}\label{f5}
f_{5}=-\alpha_{1}\beta_{1}\eta_{21}-\alpha_{0}\beta_{2}\eta
_{21}-\alpha_{1}\eta_{22}\lambda_{1}+\alpha_{0}\eta_{11}\lambda_{2}~,
\end{equation}%
\begin{equation}\label{f6}
f_{6}=-2\left(  \alpha_{2}\right)  ^{2}\eta_{21}~,
\end{equation}%
\begin{equation}\label{f7}
f_{7}=-3\alpha_{1}\alpha_{2}\eta_{21}~,
\end{equation}%
\begin{equation}\label{f8}
f_{8}=-\left(  \alpha_{1}\right)  ^{2}\eta_{21}-2\alpha_{0}%
\alpha_{2}\eta_{21}~,
\end{equation}%
\begin{equation}\label{f9}
f_{9}=-\alpha_{0}\alpha_{1}\eta_{21}~;
\end{equation}%
\begin{align}\label{g0}
g_{0}  &  =\beta_{1}\gamma\eta_{21}+\beta_{1}\eta_{22}\mu
+\beta_{1}\xi_{2}-\gamma\eta_{11}\lambda_{1}\nonumber\\
&  -\eta_{12}\lambda_{1}\mu-\lambda_{1}\xi_{1}~,
\end{align}%
\begin{align}\label{g1}
g_{1}  &  =\alpha_{1}\gamma\eta_{21}+\alpha_{1}\eta_{22}%
\mu+\alpha_{1}\xi_{2}-\beta_{2}\eta_{11}\lambda_{1}\nonumber\\
\  &  +\beta_{1}\eta_{22}\lambda_{2}+\beta_{1}\beta_{2}\eta_{21}-\eta
_{12}\lambda_{1}\lambda_{2}~,
\end{align}%
\begin{align}\label{g2}
g_{2}  &  =2\alpha_{0}\gamma\eta_{21}+2\alpha_{0}\eta_{22}%
\mu+2\alpha_{0}\xi_{2}-\beta_{1}\eta_{11}\lambda_{1}\nonumber\\
&  +\beta_{1}\eta_{22}\lambda_{1}+\left(  \beta_{1}\right)  ^{2}\eta_{21}%
-\eta_{12}\left(  \lambda_{1}\right)  ^{2}~,
\end{align}%
\begin{equation}\label{g3}
g_{3}=\alpha_{2}\beta_{1}\eta_{21}+\alpha_{1}\beta_{2}\eta
_{21}-\alpha_{2}\eta_{11}\lambda_{1}+\alpha_{1}\eta_{22}\lambda_{2}~,
\end{equation}%
\begin{align}\label{g4}
g_{4}  &  =2\alpha_{1}\beta_{1}\eta_{21}+2\alpha_{0}\beta_{2}%
\eta_{21}-\alpha_{1}\eta_{11}\lambda_{1}\nonumber\\
&  +\alpha_{1}\eta_{22}\lambda_{1}+2\alpha_{0}\eta_{22}\lambda_{2}~,
\end{align}%
\begin{equation}\label{g5}
g_{5}=3\alpha_{0}\beta_{1}\eta_{21}-\alpha_{0}\eta_{11}%
\lambda_{1}+2\alpha_{0}\eta_{22}\lambda_{1}~,
\end{equation}%
\begin{equation}\label{g6}
g_{6}=\alpha_{1}\alpha_{2}\eta_{21}~,
\end{equation}%
\begin{equation}\label{g7}
g_{7}=\left(  \alpha_{1}\right)  ^{2}\eta_{21}+2\alpha_{0}%
\alpha_{2}\eta_{21}~,
\end{equation}%
\begin{equation}\label{g8}
g_{8}=3\alpha_{0}\alpha_{1}\eta_{21}~,
\end{equation}%
\begin{equation}\label{g9}
g_{9}=2\left(  \alpha_{0}\right)  ^{2}\eta_{21}~;
\end{equation}%
\begin{equation}\label{h0}
h_{0}=\beta_{1}\lambda_{2}-\beta_{2}\lambda_{1}~,
\end{equation}%
\begin{equation}\label{h1}
h_{1}=2\alpha_{0}\lambda_{2}-\alpha_{1}\lambda_{1}~,
\end{equation}%
\begin{equation}\label{h2}
h_{2}=\alpha_{1}\lambda_{2}-2\alpha_{2}\lambda_{1}~.
\end{equation}

$12$ \textit{constraints} that must be satisfied by the $23$ \textit{coefficients}
$f_{k}$,$~g_{k}~$and $h_{\ell}$ as a consequence of the fact that they are
expressed by these explicit formulas in terms of (only) $15$ \textit{a priori arbitrary} \textit{parameters} $\eta_{nm},$ $\xi_{n},$ $\alpha_{\ell},$ $\beta_{n},$
$\gamma,$ $\lambda_{n},$ $\mu$ are reported in \textbf{Subsection 2.4}. For instance, in the case $h_1\neq0$ and $h_2\neq0$ one has the 12 constraints reported in eqs. (\ref{12Constrains a}) and (\ref{12Constrains b}), where  the $12$ \textit{coefficients} $f_{9},$ $f_{8},$ $f_{7},$ $f_{6},$ $f_{4},$ $f_{0},$ $g_{7},$ $g_{5},$ $g_{4}$, $g_{3},$ $g_{2},$ $g_{1},$ are given (by $12$
\textit{explicit} formulas) in terms of the $8$ \textit{coefficients} $f_{5},$ $f_{3},$
$g_{9,}$ $g_{8},$ $g_{6},$ $h_{2},$ $h_{1},$ $h_{0}$ , while the remaining $3$ coefficients $f_{2},$ $f_{1},$ $g_{0},$ may be \textit{freely assigned} (up to
certain minor limitations, mainly due to the way we have displayed these formulas). 
\end{subequations}

\bigskip

\subsection{The isochronous case}\label{section  isochronous case}

In this \textbf{Subsection 2.3 }we discuss the important case in which our
system (\ref{ODEs}) is \textit{isochronous}, namely \textit{all} its
\text{nonsingular} solutions $x_{n}\left(  t\right)  $ are \textit{periodic} with the
\textit{same} period $T$:
\begin{equation}
x_{n}\left(  t+T\right)  =x_{n}\left(  t\right)  ~.
\end{equation}
As already mentioned in the last part of \textbf{Subsection \ref{An auxiliary, solvable system of $2$ linear ODEs}}, this happens provided the $4$ \textit{parameters} $\eta_{nm}$ satisfy the $2$ simple \textit{restrictions} in eq. (\ref{restriction eta}) (an
\textit{equality} and an \textit{inequality}), and the period $T=2\pi/\omega$\ is then defined by the standard formula, see eqs.  (\ref{isochronous y solutions}).

Then of course the solutions $x_{n}\left(  t\right)  $ are defined by the
formulas (\ref{Solx1x2(t)})-(\ref{Solx1x2(t) 2}) with (\ref{isochronous y solutions}), where the \textit{initial} values
$y_{n}\left(  0\right)  $ are given in terms of the \textit{initial} values
$x_{n}\left(  0\right)  $ by the formulas (\ref{xyt}), giving:
\begin{subequations}
\label{y120}%
\begin{equation}
y_{1}\left(  0\right)  =\alpha_{0}\left[  x_{1}\left(  0\right)  \right]
^{2}+\alpha_{1}x_{1}\left(  0\right)  x_{2}\left(  0\right)  +\alpha
_{2}\left[  x_{2}\left(  0\right)  \right]  ^{2}+\beta_{1}x_{1}\left(
0\right)  +\beta_{2}x_{2}\left(  0\right)  +\gamma~,
\end{equation}%
\begin{equation}
y_{2}\left(  0\right)  =\lambda_{1}x_{1}\left(  0\right)
+\lambda_{2}x_{2}\left(  0\right)  +\mu~.
\end{equation}
It shall then automatically turn out that the solutions $x_{n}\left(
t\right)  $ are \textit{real}---provided of course that the \textit{initial
values} $x_{n}\left(  0\right)  $ are themselves \textit{real}, as well as all
the other \textit{parameters} and \textit{coefficients}---as long as they are well defined. Of
course, the solutions of \textit{nonlinear} systems of ODEs may at some time
\textit{blow-up} or become \textit{undetermined}; remarkably, for our
solutions, the \textit{first} phenomenon does \textit{not} happen; while the \textit{second} of
course may happen since the sign to be attributed to the square-root in the
right-hand side of eq. (\ref{S(t)}) is defined at $t=0$ (by requiring the
consistency of the formulas (\ref{Solx1x2(t)}) and (\ref{y120}) at $t=0$), but
this determination is no more well defined for $t>t_S$ if at some (\textit{finite, positive}) value
$t_{S}$  the quantity $S\left(  t_{S}\right)  $ vanishes:
\end{subequations}
\begin{equation}
S\left(  t_{S}\right)  =0~.
\end{equation}

Examples of solutions of the system (\ref{ODEs})---both in the \textit{isochronous} and \textit{not isochronous} cases---are presented in \textbf{Section 3}.

\bigskip

\subsection{The constraints on the coefficients of the system of ODEs (\ref{ODEs})}\label{Section constraints}

In this \textbf{Subsection 2.4} we show how to invert the relations (\ref{f1f2g}) and report the \textit{constraints} mentioned in \textbf{Subsection 2.2}. We will describe how to achieve  this goal for different values of $h_1$ and $h_2$ separately.

We emphasize that the existence of $12$ \textit{constraints} on the $23$ \textit{coefficients} $f_{k}%
$,$~g_{k}~$and $h_{\ell}$ implies that $4$ of the $15$ \textit{parameters} $\eta_{nm},$
$\xi_{n},$ $\alpha_{\ell},$ $\beta_{n},$ $\gamma,$ $\lambda_{n},$ $\mu$ are in
some sense \textit{redundant}. In fact, due to the existence of these 12 \textit{constraints}, \textit{only} $11$ of the $23$ coefficients $f_{k}$, $~g_{k}~$ and $h_{\ell}$ can be assigned \textit{arbitrarily}. Indeed, one can invert the expression of these $11$ \textit{independent coefficients} in terms of (only!) $11$ of the \textit{parameters} $\eta_{nm},$ $\xi_{n},$ $\alpha_{\ell},$ $\beta_{n},$ $\gamma,$ $\lambda_{n},$ $\mu$, using eqs. (\ref{f1f2g}).

Note that the inversion procedure described in the next subsections will leave the 4 \textit{parameters} $\beta_{2}$,
$\xi_{2}$, $\gamma$, and $\eta_{22}$ \textit{undetermined}.
\footnote{However, this choice is arbitrary, and one may choose another set of 4 arbitrary parameters. For instance, in the example treated in Subsection \ref{Section Example 1} we will make a different choice, assigning arbitrary values to $\beta_{2}$, $\gamma$, $\eta_{22}$ and $\mu$.} This means that, starting from the system of ODEs (\ref{ODEs}) with given \textit{coefficients} $f_{k},~g_{k}%
~$and $h_{\ell}$ satisfying the 12 \textit{constraints} mentioned above (and reported below), one finds a \textit{linear} system (\ref{ydot}) which depends on 4 \textit{arbitrary} \textit{parameters} $\beta_{2}$,
$\xi_{2}$, $\gamma$, and $\eta_{22}$. However, although the solution of
(\ref{ydot}) depends on these 4 \textit{arbitrary} \textit{parameters}, the change of
variables (\ref{Solx1x2(t)}) depends on these \textit{parameters} in such a way
that the final dependence of the solutions of eqs. (\ref{ODEs}) on these
4 \textit{parameters} $\beta_{2}$, $\xi_{2}$, $\gamma$, and $\eta_{22}$ is eliminated.

\subsubsection{Case $h_2\neq 0$ and  $h_1\neq 0$.}

Let us assume that both the \textit{parameters} $h_{1}$ and $h_{2}$ are \textit{nonzero}. In this case, we can invert the
$3$ eqs. (\ref{h0}), (\ref{h1}), and (\ref{h2}) in terms of $\lambda_{1}%
$, $\lambda_{2}$, and $\beta_{1}$, obtaining
\begin{subequations}
\label{inversion lambda beta1}%
\begin{equation}
\lambda_{1}=\left(  \alpha_{1}h_{1}-2\alpha_{0}h_{2}\right)  /\left[
4\alpha_{0}\alpha_{2}-\left(  \alpha_{1}\right)  ^{2}\right]  \,,
\label{landa1}%
\end{equation}

\begin{equation}
\lambda_{2}=\left(  \alpha_{1}h_{2}-2\alpha
_{2}h_{1}\right)  /\left[  \left(  \alpha_{1}\right)  ^{2}-4\alpha_{0}%
\alpha_{2}\right]  \,, \label{landas}%
\end{equation}

\begin{equation}
\beta_{1}=\left[  -\alpha_{1}\beta_{2}%
h_{1}+2\alpha_{0}\beta_{2}h_{2}+\left(  \alpha_{1}\right)  ^{2}h_{0}%
-4\alpha_{0}\alpha_{2}h_{0}\right]  /\left(  \alpha_{1}h_{2}-2\alpha_{2}%
h_{1}\right)  \,. \label{beta1}%
\end{equation}
We then insert eqs. (\ref{inversion lambda beta1}) into eqs. (\ref{f1f2g}),
and solve eqs. \ (\ref{g6}), (\ref{g8}), and (\ref{g9}) for $\alpha_{1}$,
$\alpha_{2}$ and $\eta_{21}$, obtaining%
\end{subequations}
\begin{equation}
\alpha_{1}=2\alpha_{0}g_{8}/\left(  3g_{9}\right)  ,\,\,\alpha_{2}=3\alpha
_{0}g_{6}/g_{8},\,\,\eta_{21}=g_{9}/\left[  2\left(  \alpha_{0}\right)
^{2}\right]  \,. \label{inversion alpha1 alpha2  eta21}%
\end{equation}
Then, we insert eqs. (\ref{inversion alpha1 alpha2 eta21}) into eqs. (\ref{f1f2g}),
and solve eq. (\ref{f5}) for $\alpha_{0}$, getting
\begin{align}
&  \alpha_{0}=3\beta_{2}g_{6}g_{9}\left(  27g_{6}\left(  g_{9}\right)
^{2}-\left(  g_{8}\right)  ^{3}\right)  h_{1}/\left\{  27\left(  g_{6}%
g_{9}\right)  ^{2}\left[  3h_{1}\left(  f_{5}-\eta_{22}h_{1}\right)
+2g_{8}h_{0}\right]  \right. \nonumber\\
&  \left.  +3f_{3}g_{9}\left(  g_{8}\right)  ^{3}h_{2}-g_{6}g_{8}\left[
9g_{9}g_{8}h_{2}\left(  f_{5}-\eta_{22}h_{1}\right)  +27f_{3}\left(
g_{9}\right)  ^{2}h_{1}+2\left(  g_{8}\right)  ^{3}h_{0}\right]  \right\}  \,.
\label{inversion alpha0}%
\end{align}
Subsequently, we insert eq. (\ref{inversion alpha0}) into eqs. (\ref{f1f2g}), and
solve eq. (\ref{f3}) for $\eta_{11}$, getting
\begin{align}
&  \eta_{11}=\left\{  81\left(  g_{6}g_{9}\right)  ^{2}\left[  h_{1}\left(
3f_{5}-\eta_{22}h_{1}\right)  +2g_{8}h_{0}\right]  +f_{3}\left(  g_{8}\right)
^{3}\left(  9g_{9}h_{2}-2g_{8}h_{1}\right)  \right. \nonumber\\
&  \left.  -3g_{6}g_{8}\left[  3g_{9}g_{8}h_{2}\left(  3f_{5}-\eta_{22}%
h_{1}\right)  +9f_{3}\left(  g_{9}\right)  ^{2}h_{1}+2\left(  g_{8}\right)
^{3}h_{0}\right]  \right\} \nonumber\\
&  /\left[  9g_{6}g_{9}h_{1}\left(  9g_{6}g_{9}h_{1}-\left(  g_{8}\right)
^{2}h_{2}\right)  \right]  \,. \label{inversion eta11}%
\end{align}

By inserting this expression (\ref{inversion eta11}) of $\eta_{11}$ in the
eqs. (\ref{f1f2g}), we note that eqs. (\ref{f4}), (\ref{f6}), (\ref{f7}), (\ref{f8}), (\ref{f9}), (\ref{g3}), (\ref{g4}), (\ref{g5}), (\ref{g7}) do not feature the $15$ \textit{parameters} $\eta_{nm},$ $\xi_{n},$ $\alpha_{\ell},$
$\beta_{n},$ $\gamma,$ $\lambda_{n},$ $\mu$, indeed they give the following
$9$ \textit{constraints}: 
\begin{subequations}
\label{12Constrains a}%
\begin{equation}
f_{9}=-g_{8}/3~, \label{vf9}%
\end{equation}%
\begin{equation}
f_{8}=-2\left(  g_{8}\right)  ^{2}/\left(
9g_{9}\right)  -3g_{6}g_{9}/g_{8}~, \label{vf8}%
\end{equation}%
\begin{equation}
f_{7}=-3g_{6}~, \label{vf7}%
\end{equation}%
\begin{equation}
f_{6}=-9\left(  g_{6}\right)  ^{2}g_{9}/\left(
g_{8}\right)  ^{2}~, \label{vf6}%
\end{equation}%
\begin{align}
f_{4}  &  =\left\{  f_{3}\left(  g_{8}\right)
^{2}\left(  2g_{8}h_{1}-3g_{9}h_{2}\right)  +g_{6}\left[  9f_{5}g_{9}%
g_{8}h_{2}+2\left(  g_{8}\right)  ^{3}h_{0}\right]  \right. \nonumber\\
&  -54\left(  g_{6}g_{9}\right)  ^{2}h_{0}\left.  {}\right\}  /\left(
9g_{6}g_{8}g_{9}h_{1}\right)  ~, \label{vf4}%
\end{align}

\begin{equation}
g_{7}=2\left(  g_{8}\right)  ^{2}/\left(  9g_{9}\right)
+3g_{6}g_{9}/g_{8}=-f_{8}~, \label{vg7}%
\end{equation}%
\begin{align}
&  g_{5}=\left[  f_{3}\left(  g_{8}\right)  ^{2}\left(
3g_{9}h_{2}-g_{8}h_{1}\right)  -3g_{6}\left(  g_{8}\right)  ^{3}%
h_{0}+81\left(  g_{6}g_{9}\right)  ^{2}h_{0}\right]
\nonumber\\
&  /\left\{  3g_{6}\left[  9g_{6}g_{9}h_{1}-\left(  g_{8}\right)  ^{2}%
h_{2}\right]  \right\}  ~, \label{vg5}%
\end{align}%
\begin{align}
&  g_{4}=\left\{  g_{8}\left[  2\left(  \left(
g_{8}\right)  ^{3}-27g_{6}\left(  g_{9}\right)  ^{2}\right)  \left(
f_{3}h_{1}+3g_{6}h_{0}\right)  \right]  /\left[  g_{9}\left(  \left(
g_{8}\right)  ^{2}h_{2}-9g_{6}g_{9}h_{1}\right)  \right]  \right.
\nonumber\\
&  \left.  -9f_{3}g_{8}\right\}  /\left(  9g_{6}\right)  +f_{5}~, \label{vg4}%
\end{align}

\begin{align}
&  g_{3}=\left[  \left(  \left(  g_{8}\right)
^{3}-27g_{6}\left(  g_{9}\right)  ^{2}\right)  \left(  f_{3}h_{1}+3g_{6}%
h_{0}\right)  \right]  /\left[  \left(  g_{8}\right)  ^{3}h_{2}-9g_{6}%
g_{8}g_{9}h_{1}\right] \nonumber\\
&  +\left[  2\left(  g_{8}\right)  ^{3}h_{0}-27\left(  g_{9}\right)
^{2}\left(  f_{3}h_{1}+2g_{6}h_{0}\right)  \right]  /\left[  9g_{8}g_{9}%
h_{1}\right] \nonumber\\
&  +\left[  h_{2}\left(  3f_{5}g_{6}-f_{3}g_{8}\right)  \right]  /\left[
3g_{6}h_{1}\right]  ~. \label{vg3}%
\end{align}

Subsequently, we solve eq. (\ref{f2}) for $\mu$. We then insert that expression of $\mu$ in eq. (\ref{f1}) and solve this equation for $\eta_{12}$. Next, we insert the values of $\eta_{12}$
and $\mu$ into eq. (\ref{g0}), and solve for $\xi_{1}$ (we do not show the relevant formulas here, as they are too long); and by re-inserting
these findings in the $3$ eqs. (\ref{f0}), (\ref{g1}), (\ref{g2}) we finally obtain the following $3$ \textit{constraints}:
\end{subequations}
\begin{subequations}
\label{12Constrains b}%
\begin{align}
&  f_{0}=\left\{  27\left(  g_{6}g_{9}\right)  ^{2}\left[  -3\left(
h_{1}\right)  ^{2}\left(  f_{2}h_{0}+g_{0}h_{2}\right)  +6f_{5}h_{1}\left(
h_{0}\right)  ^{2}+4g_{8}\left(  h_{0}\right)  ^{3}\right]  \right.
\nonumber\\
&  -2f_{3}\left(  g_{8}\right)  ^{3}\left(  h_{0}\right)  ^{2}\left(
g_{8}h_{1}-3g_{9}h_{2}\right)  +g_{6}g_{8}\left[  -4\left(  g_{8}h_{0}\right)
^{3}-27f_{1}\left(  g_{9}\right)  ^{2}h_{1}h_{2}h_{0}\right. \nonumber\\
&  +9g_{8}g_{9}\left\{  \left(  h_{2}\left[  h_{1}\left(  f_{2}h_{0}%
+g_{0}h_{2}\right)  -2f_{5}\left(  h_{0}\right)  ^{2}\right]  +f_{1}%
h_{0}\left(  h_{1}\right)  ^{2}\right)  \right\}  \left.  {}\right]  \left.
{}\right\} \nonumber\\
&  /\left[  9g_{6}g_{8}g_{9}h_{1}h_{2}\left(  g_{8}h_{1}-3g_{9}h_{2}\right)
\right]  ~, \label{vf0}%
\end{align}

\begin{align}
&  g_{2}=\left\{  -f_{3}\left(  g_{8}\right)  ^{2}%
h_{0}\left(  2g_{8}h_{1}+3g_{9}h_{2}\right)  \left(  g_{8}h_{1}-3g_{9}%
h_{2}\right)  +g_{6}g_{8}\left\{  {}\right.  9f_{1}g_{9}g_{8}\left(
h_{1}\right)  ^{3}\right. \nonumber\\
&  \left.  +27\left(  g_{9}h_{2}\right)  ^{2}\left(  f_{2}h_{1}-f_{5}%
h_{0}\right)  -3g_{9}h_{2}\left[  3f_{5}g_{8}h_{1}h_{0}+9f_{1}g_{9}\left(
h_{1}\right)  ^{2}+2\left(  g_{8}h_{0}\right)  ^{2}\right]  \right.
\nonumber\\
&    -4\left(  g_{8}\right)  ^{3}\left(  h_{0}\right)  ^{2}h_{1}\left.
{}\right\}  +27\left(  g_{6}g_{9}\right)  ^{2}\left[  -3f_{2}\left(
h_{1}\right)  ^{3}+6f_{5}h_{0}\left(  h_{1}\right)  ^{2}+4g_{8}\left(
h_{0}\right)  ^{2}h_{1}\right.\nonumber\\
&\left.\left. +6g_{9}\left(  h_{0}\right)  ^{2}h_{2}\right]  \right\}
  /\left\{  9g_{6}g_{9}h_{1}h_{2}\left[  9g_{6}g_{9}h_{1}-\left(
g_{8}\right)  ^{2}h_{2}\right]  \right\}  ~, \label{vg2}%
\end{align}

\begin{align}\label{vg1}
&  g_{1}=\left\{  27f_{5}g_{6}h_{0}-9\left(  f_{3}%
g_{8}h_{0}+f_{2}g_{6}h_{1}\right)  +\right. \nonumber\\
&  \left.  \left[  2g_{8}\left(  \left(  g_{8}\right)  ^{3}-27g_{6}\left(
g_{9}\right)  ^{2}\right)  h_{0}\left(  f_{3}h_{1}+3g_{6}h_{0}\right)
\right]  /\left[  g_{9}\left(  \left(  g_{8}\right)  ^{2}h_{2}-9g_{6}%
g_{9}h_{1}\right)  \right]  \right\} \nonumber\\
&  /\left(  9g_{6}h_{1}\right)  ~.
\end{align}

So, in conclusion, for $h_{1}\neq0$ and $h_{2}\neq0$, the $12$ \textit{constraints} given by
eqs. (\ref{12Constrains a}) and (\ref{12Constrains b}) on the $23$
\textit{coefficients} $f_{k},~g_{k}~$and $h_{\ell}$ express the $12$ \textit{coefficients}
$f_{9},$ $f_{8},$ $f_{7},$ $f_{6},$ $f_{4},$ $f_{0},$ $g_{7},$ $g_{5},$
$g_{3}$ $g_{4},$ $g_{2},$ $g_{1},$ in terms of (only!) the $8$ \textit{coefficients}
$f_{5},$ $f_{3},$ $g_{9,}$ $g_{8},$ $g_{6},$ $h_{2},$ $h_{1},$ $h_{0}$ (which
may be \textit{generically} assigned, as well as the remaining 3 \textit{coefficients}
$f_{1},$ $f_{2},$ $g_{0}$; except for the obvious requirement that the
previous equations make sense, hence that their denominators do not vanish).

\subsubsection{Case $h_2=0$,  $h_1\neq 0$ (or $h_2\neq0$,  $h_1= 0$).}

The existence of $12$ \textit{constraints} between the $23$ \textit{coefficients} $f_{k}%
,~g_{k}~$and $h_{\ell}$ occurs generically, and it is not restricted to the case $h_{2}\neq0$, $h_{1}\neq0$ treated in the preceding \textbf{Subsubsection 2.4.1}. If instead $h_{1}\neq0$ and $h_{2}=0$, we can proceed as above (in \textbf{Subsubsection 2.4.1}), until eqs. (\ref{12Constrains a}). Then, we can again
solve eq. (\ref{f2}) for $\mu$, but in this case, by inserting the expression thereby obtained
for $\mu$ into eq. (\ref{f1}), we obtain an equation that does not feature the\textit{ parameters} $\eta_{12}$, $\beta_{2}$, $\xi_{1}$, $\xi_{2}$, and $\gamma
$, so that this equation gives a \textit{constraint} between the \textit{coefficients} $f_{k}$,
$g_{k}$, and $h_{\ell}$. Then, we must insert the expression of $\mu$ in eq.
(\ref{g2}) and solve this equation for $\eta_{12}$. Next, we insert the values of $\eta_{12}$ and $\mu$ into eqs. (\ref{f1f2g}), and solve   eq. (\ref{g0}) for $\xi_{1}$.
Finally, by re-inserting these findings in the $3$ eqs. (\ref{f0}), (\ref{f1})
and (\ref{g1}), we obtain
\end{subequations}
\begin{subequations}\label{12Constrains c}
\begin{align}
&  f_{0}=\left\{  {}\right.  g_{8}\left(  h_{0}\right)  ^{2}\left(  f_{3}%
h_{1}/g_{6}+2h_{0}\right)  +3h_{1}h_{0}\left(  f_{2}h_{1}-f_{5}h_{0}\right) \nonumber\\
&  -27g_{6}g_{9}\left[  2g_{9}\left(  h_{0}\right)  ^{3}+\left(  h_{1}\right)
^{2}\left(  g_{0}h_{1}-g_{2}h_{0}\right)  \right]  /\left(  g_{8}\right)
^{2}\left.  {}\right\}  /\left[  3\left(  h_{1}\right)  ^{3}\right]  ~,
\end{align}

\begin{align}
&f_{1}= \left\{2\left(  g_{8}\right)  ^{4}\left[  f_{3}h_{0}h_{1}+2g_{6}\left(
h_{0}\right)  ^{2}\right] \right. \nonumber\\
&\left.  -27\left(  g_{6}g_{9}\right)  ^{2}\left[
6f_{5}h_{1}h_{0}-3f_{2}\left(  h_{1}\right)  ^{2}+4g_{8}\left(  h_{0}\right)^{2}\right]  \right\}
/\left[9g_{6}\left(  g_{8}\right)  ^{2}g_{9}\left(  h_{1}\right)
^{2}\right]\,,
\end{align}

\begin{align}
&g_{1}=\left\{-2f_{3}\left(  g_{8}\right)  ^{4}h_{0}h_{1}-3g_{6}g_{8}%
h_{0}\left[  9f_{3}\left(  g_{9}\right)  ^{2}h_{1}+2\left(  g_{8}\right)
^{3}h_{0}\right] \right. \nonumber\\
& \left. +81\left(  g_{6}g_{9}\right)  ^{2}\left[  h_{1}\left(
3f_{5}h_{0}-f_{2}h_{1}\right)  +2g_{8}\left(  h_{0}\right)  ^{2}\right] \right\}
/{\left(  9g_{6}g_{9}h_{1}\right)  ^{2}}~.
\end{align}
\end{subequations}
We therefore conclude that, in this case $h_{1}\neq0$ and $h_{2}=0$, the $12$ \textit{constraints} on the \textit{coefficients} $f_{k}$, $g_{k}$, $h_{1}$ and $h_{0}$  are
given by eqs. (\ref{12Constrains a}) and (\ref{12Constrains c}).

The case $h_2\neq0$,  $h_1= 0$ is simply obtained by exchanging $x_1$ with $x_2$, which of course corresponds to a renumbering of the coefficients $f_k$, $g_k$ and $h_l$ in eqs. (\ref{12Constrains a}) and (\ref{12Constrains c}).

\subsubsection{Case $h_{1}=h_{2}=0$ and $h_{0}=1$.}\label{case h10= h2=0}

In the case $h_{1}=h_{2}=0$ we start by solving the eqs. (\ref{h1}) and (\ref{h2}) for $h_{1}=h_{2}=0$ for $\lambda_{1}$ and $\lambda_{2}$. If $\left(  \alpha_{1}\right)
^{2}\neq4\alpha_{0}\alpha_{2}$, one gets $\lambda_{1}=\lambda_{2}=0$, which
also implies that $h_{0}=0$, see eq. (\ref{h0}), hence this case is \textit{not interesting} (see also eq. (\ref{xyt})). We therefore assume $\left(
\alpha_{1}\right)  ^{2}=4\alpha_{0}\alpha_{2}$, setting%
\begin{equation}
\alpha_{0}=\left(  \alpha_{1}\right)  ^{2}/(4\alpha_{2})\,, \label{alpha0 2}%
\end{equation}
assuming $\alpha_{2}\neq0$. Note that in this case eqs. (\ref{h1}) and
(\ref{h2}) are the same, so we can ignore one of them and also set 
 $h_{0}=1$ without any loss of generality.

Solving eq. (\ref{h0}) and eq. (\ref{h1}) for $\lambda_{1}$ and $\lambda_{2}$ we get
\begin{subequations}
\label{inversion lambda beta2}%
\begin{equation}
\lambda_{1}=-\alpha_{1}/\left(  \alpha_{1}\beta_{2}-2\alpha_{2}\beta
_{1}\right)  ,
\end{equation}
\begin{equation}
\lambda_{2}=2\alpha_{2}/\left(  2\alpha
_{2}\beta_{1}-\alpha_{1}\beta_{2}\right)  ~. %
\end{equation}
We then insert eqs. (\ref{alpha0 2}) and (\ref{inversion lambda beta2}) into eqs. (\ref{f1f2g}),
and solve eqs. \ (\ref{g9}), (\ref{g6}) for $\alpha_{1}$ and $\eta_{21}$,
obtaining
\end{subequations}
\begin{equation}
\alpha_{1}=2\alpha_{2}\left(  g_{9}/g_{6}\right)  ^{1/3}~,~~~\eta_{21}=\left(
g_{6}\right)  ^{4/3}/\left[  2\left(  \alpha_{2}\right)  ^{2}\left(
g_{9}\right)  ^{1/3}\right]  \,. \label{inversion alpha1 eta12}%
\end{equation}

Next, we insert eqs. (\ref{inversion alpha1 eta12}) into eqs. (\ref{f1f2g}) and solve
eq. (\ref{f5}) for $\eta_{11}$, getting%
\begin{align}
&  \eta_{11}=2\eta_{22}+\left(  g_{6}\right)  ^{1/3}\left\{  2\alpha_{2}%
f_{5}+\beta_{2}\left[  \left(  g_{6}\right)  ^{2}{g_{9}}\right]  ^{1/3}%
+2\beta_{1}g_{6}\right\}  \left[  \beta_{1}\left(  {g_{6}}\right)
^{1/3}-\beta_{2}\left(  {g_{9}}\right)  ^{1/3}\right]
\label{inversion eta11 2}\\
&  /\left\{  2\left[  \alpha_{2}\left(  g_{9}\right)  ^{1/3}\right]  ^{2}%
\right\}  \,.\nonumber
\end{align}
By inserting eqs. (\ref{inversion eta11 2}) into eqs. (\ref{f1f2g}), and solving eq. (\ref{f3})
for $\alpha_{2}$ we get%
\begin{align}\label{inversion alpha2 2}
\alpha_{2}=\left\{  \beta_{2}\left[  \left(  g_{6}\right)  ^{4}{g_{9}}\right]
^{1/3}-\beta_{1}\left(  g_{6}\right)  ^{5/3}\right\}  /\left[  f_{5}\left(
g_{6}\right)  ^{2/3}-f_{3}\left(  {g_{9}}\right)  ^{2/3}\right]  \,.
\end{align}
We then insert eq. (\ref{inversion alpha2 2}) into eq. (\ref{f1f2g}), solve
eqs. (\ref{f2}) for $\xi_{2}$, and then insert the result into eqs. (\ref{f1f2g}). We then solve
eq. (\ref{f1}) for $\beta_1$, and  insert the result into eqs. (\ref{f1f2g}). Finally, we solve
eqs. (\ref{f0}) and (\ref{g0}) for $\xi_{1}$ and $\eta_{12}$. Inserting the resulting expressions into eqs. (\ref{f1f2g})
 we find that the remaining 12 equations---namely eqs. (\ref{f9}), (\ref{f8}), (\ref{f7}), (\ref{f6}), (\ref{f4}), (\ref{g8}), (\ref{g7}), (\ref{g5}), (\ref{g3}), (\ref{g4}), (\ref{g2}), (\ref{g1})---do not depend on the 15 \textit{parameters}
$\eta_{nm},$ $\xi_{n},$ $\alpha_{\ell},$ $\beta_{n},$ $\gamma,$ $\lambda_{n},$
$\mu$, so that they are 12 \textit{constraints} on the 23 \textit{coefficients} $f_{k}$,$~g_{k}%
~$and $h_{\ell}$, which then read as follows:
\begin{subequations}
\label{12Constrains 2}%
\begin{equation}
f_{9}=-\left[  g_{6}\left(  g_{9}\right)  ^{2}\right]  ^{1/3}\,,
\end{equation}

\begin{equation}
f_{8}=-3\left[  \left(  g_{6}\right)  ^{2}g_{9}\right]
^{1/3}\,, %
\end{equation}

\begin{equation}
f_{7}=-3 g_{6} \, ,
\end{equation}

\begin{equation}
f_{6}=-\left[  \left(  g_{6}\right)  ^{4}/{g_{9}%
}\right]  ^{1/3}\,, %
\end{equation}

\begin{equation}
f_{4}=f_{5}\left(  {g_{6}}/g_{9}\right)  ^{1/3}%
+f_{3}\left(  {g_{9}}/g_{6}\right)  ^{1/3}\,, 
\end{equation}

\begin{equation}
g_{8}=3\,\left[  g_{6}\left(  g_{9}\right)  ^{2}\right]
^{1/3}=-3\,f_{9}~, 
\end{equation}

\begin{equation}
g_{7}=3\,\left[  \left(  g_{6}\right)  ^{2}g_{9}\right]
^{1/3}=-f_{8}\,, 
\end{equation}

\begin{equation}
g_{5}=\left\{  f_{3}g_{9}-3f_{5}\left[  \left(
g_{6}\right)  ^{2}{g_{9}}\right]  ^{1/3}\right\}  /\left(  2g_{6}\right)  \,,
\end{equation}

\begin{equation}
g_{3}=-\left[  f_{5}\left(  g_{6}\right)  ^{2/3}%
+f_{3}\left(  g_{9}\right)  ^{2/3}\right]  /\left[  2\left(  {g_{6}g_{9}%
}\right)  ^{1/3}\right]  \,, 
\end{equation}

\begin{equation}
g_{4}=-2f_{5}\,,
\end{equation}

\begin{equation}
g_{2}=\left\{  \left[  f_{5}\left(  g_{6}\right)
^{1/3}\right]  ^{2}-f_{1}g_{9}\left(  g_{6}\right)  ^{2/3}-\left(
g_{9}\right)  ^{2/3}\left(  f_{2}g_{6}+f_{3}f_{5}\right)  \right\}  /\left\{
2\left[  \left(  g_{6}\right)  ^{4}\left(  {g_{9}}\right)  \right]
^{1/3}\right\}  \,, 
\end{equation}

\begin{equation}
g_{1}=\left[  -3f_{1}\left(  g_{9}/g_{6}\right)
^{1/3}/2- \left[  f_{3}\left(  g_{9}\right)
^{1/3}\right]  ^{2}/\left(  2g_{6}^{5/3}\right)
+f_{3}f_{5}/\left(  2g_{6}\right)  +f_{2}/2\right] \, . \end{equation}
\end{subequations}

\subsection{Additional properties of the subclass of the system of ODEs (\ref{ODEs}) that we identify as explicitly solvable}\label{Section additional properties}

In this  \textbf{Subsection} we sketch a terse analysis of some features of the solutions of the system of $2$ nonlinear
ODEs (\ref{ODEs}) in the \textit{explicitly solvable} case in which the $23$ coefficients $f_{k},~g_{k}~$and $h_{\ell}$ are expressed as detailed in
\textbf{Section \ref{Section main results}}.

\textbf{Remark 2.5-1.} It is convenient, before getting into the display of several examples, to introduce a significant simplification of the system of ODEs (1). It can be easily shown that eqs. (\ref{f1f2g}) imply the
following formula, 
which entails the introduction of the new polynomial of second order $P_{2}(x_{1},x_{2})$:
\begin{align}
	&  \left[  \lambda_{1}P_{3}^{\left(  1\right)  }\left(  x_{1},x_{2}\right)
	+\lambda_{2}P_{3}^{\left(  2\right)  }\left(  x_{1},x_{2}\right)  \right]
	/P_{1}\left(  x_{1},x_{2}\right)  =\nonumber\label{constraint Ps}\\
	&  \gamma\eta_{21}+\xi_{2}+\eta_{22}\mu+\alpha_{0}\eta_{21}\left(
	x_{1}\right)  ^{2}+\alpha_{1}\eta_{21}x_{2}x_{1}+\alpha_{2}\eta_{21}\left(
	x_{2}\right)  ^{2}+\beta_{1}\eta_{21}x_{1}+\beta_{2}\eta_{21}x_{2}\nonumber\\
	&  +\eta_{22}\lambda_{1}x_{1}+\eta_{22}\lambda_{2}x_{2}\equiv P_{2}%
	(x_{1},x_{2})\,.
\end{align}
This formula clearly implies that the \textit{linear} change of variables
\begin{equation}\label{linearchangeofvariables z}
	z_{1}=\left(\lambda_{2}x_{1}-\lambda_{1}x_{2}\right)/\lambda~ ,\quad z_{2}=\left(\lambda_{1}x_{1}%
	+\lambda_{2}x_{2}\right)/\lambda,\quad   \lambda\equiv \sqrt{\lambda_{1}^2+\lambda_{2}^2}
\end{equation}
causes the right-hand side of the ODE expressing $\dot{z}_{2}$ in terms of the
\textit{new} variables $z_{n}\left(  t\right)  ~$to be simply \textit{polynomial} in
these variables (no denominator!):%

\begin{subequations}\label{ODEs 2}
	\begin{equation}
		\dot{z}_{1}= \tilde P_{3}^{\left(  1\right)}(z_{1},z_{2})/\tilde P_{1}(z_{1},z_{2})\,,
	\end{equation}
	
\begin{equation}
	\dot{z}_{2}= \tilde P_{2}(z_{1},z_{2})\,,
\end{equation}
\end{subequations}
with
\begin{equation}\label{linearchangeofvariables x}
	{x}_{1}(z_{1},z_{2})=\left(\lambda_{2}z_{1}+\lambda_{1}z_{2}\right)/\lambda~,\quad {x}_{2}(z_{1},z_{2})=\left(-\lambda_{1}z_{1}+\lambda_{2}z_{2}\right)/\lambda\,,
\end{equation}
and

\begin{subequations}

\begin{equation}
\tilde P_{3}^{\left(  1\right)}(z_{1},z_{2})= \lambda_{2}P_{3}^{\left(  1\right)}[{x}_{1}(z_{1},z_{2}),{x}_{2}(z_{1},z_{2})]
	-\lambda_{1}P_{3}^{\left(  2\right)  }[{x}_{1}(z_{1},z_{2}),{x}_{2}(z_{1},z_{2})]   \,,
\end{equation}
 \begin{equation}
 	\tilde P_{2}(z_{1},z_{2}) \equiv P_{2}[{x}_{1}(z_{1},z_{2}),{x}_{2}(z_{1},z_{2})] \,,
 \end{equation}
 
\begin{equation}
	\tilde P_{1}^{\left(  1\right)}(z_{1},z_{2})= P_{1}^{\left(  1\right)}[{x}_{1}(z_{1},z_{2}),{x}_{2}(z_{1},z_{2})]  \,,
\end{equation}

\end{subequations}

This shows that the \text{subsystem} of the system of ODEs (\ref{ODEs}) on which we
focus in this paper---namely, that characterized by the simple change of
variables (\ref{xyt}), (\ref{Solx1x2(t)}), (\ref{Solx1x2(t) 2}) with the variable $y_{n}\left(  t\right)
$ satisfying the \textit{linear} system of ODEs (\ref{ydot})---may \textit{always} be reduced by the
simple \textit{linear} transformation (\ref{linearchangeofvariables x}) to an \textit{equivalent} system of ODEs
(\ref{ODEs 2}) featuring \textit{only} $10+3+6=19$ \textit{coefficients} satisfying 8 \textit{constraints}.

{
	
In the \textit{examples} in  \textbf{Subsections 3.2-3.5}, we will place the \textit{singularity line} (in the
$x_{1}-x_{2}$ Cartesian plane) of the ODEs (\ref{ODEs}), that is the line
$P_{1}(x_{1},x_{2})=0$, on the $x_{2}$ axis, setting $h_{0}=h_{2}=0$. This, of course, shall merely amount to making an appropriate translation and rotation in the $x_1-x_2$ Cartesian plane, which is convenient to simplify the presentation and display of the numerical examples. Moreover, we set $\lambda_{1}=0$, so that we will have
\begin{equation}
	\dot{x}_{2}=P_{3}^{(2)}(x_{1},x_{2})/P_{1}(x_{1},x_{2})=P_{2}(x_{1},x_{2})\,,
\end{equation}
see below \textbf{Subsections 3.2-3.5}.  But this minor simplification introduces a certain asymmetry among the 2 ODEs (1).  

}

\textbf{Remark 2.5-2.} Let us mention that whenever in the following we mention "points", e. g. $\boldsymbol{x}= \left(x_1,x_2\right)$
(or, say, $\boldsymbol{y}= \left(y_1,y_2\right)$), we always mean \textit{points} in the $x_{1}-x_{2}$
Cartesian plane (or the corresponding $y_{1}-y_{2}$ Cartesian plane); or of
course the\ \textit{values} of the corresponding \textit{numbers}.
$\blacksquare$

Clearly the \textit{equilibrium} \textit{solutions} of the \textit{linear}
system of ODEs (\ref{ydot}) are%
\begin{equation}
\boldsymbol{y}\left(  t\right) = \left(y_1(t),y_2(t)\right)=\left(u_1,u_2\right)\equiv \bar{\boldsymbol{y}} ~\,,\label{equilibrium y}%
\end{equation}
where the numbers $u_{n}$ are defined by the eqs. (\ref{u1})-(\ref{u2}). The
corresponding expressions of the \textit{equilibria} of the system of ODEs
(\ref{ODEs}) ---namely, the points $\bar{\boldsymbol{x}}=\left(\bar{x}_1,\bar{x}_2\right)  $ such
that the right-hand sides of the $2$ ODEs (\ref{ODEs}) \textit{vanish}---are
then given by the eqs. (\ref{Solx1x2(t)}) with $y_{n}\left(  t\right)  =u_{n}$, namely

\begin{subequations}
\label{Equilibrium points}%
\begin{equation}
	\bar{\boldsymbol{x}}^{(+)}=\left(  X_{1}^{(+)}\left(  u_{1},u_{2}\right)  ,X_{2}^{\left(
		+\right)  }\left(  u_{1},u_{2}\right)  \right)   \,,
\end{equation}

\begin{equation}
	\bar{\boldsymbol{x}}^{(-)}=\left(  X_{1}^{(-)}\left(
	u_{1},u_{2}\right)  ,X_{2}^{\left(  -\right)  }\left(  u_{1},u_{2}\right)
	\right)  \,.
\end{equation}
\end{subequations}
If $\boldsymbol{\bar{x}}^{(+)  }$ and $\boldsymbol{\bar{x}}^{(-)  }$ are \textit{complex}, which happens when the
quantity $S(t)$ defined by eq. (\ref{S(t)}) is \textit{complex} when evaluated
for $y_{n}(t)=u_{n}$, the system of ODEs (\ref{ODEs}) has \textit{no real
equilibrium} \textit{solution}. When $\boldsymbol{\bar{x}}^{(+)  }$ and $\boldsymbol{\bar{x}}^{(-)  }$ are
instead \textit{real}, which happens when the quantity $S(t)$ defined by eq.
(\ref{S(t)}) is \textit{real} when evaluated for $y_{n}(t)=u_{n}$, the system
of ODEs (\ref{ODEs}) has $2$ \textit{distinct} \textit{real}
\textit{equilibrium} \textit{solutions}.

When $\boldsymbol{\bar{x}}^{(+)  }$ and $\boldsymbol{\bar{x}}^{(-)  }$ 
are \textit{real} and \textit{distinct}, these \textit{equilibria} may be
\textit{stable equilibria} (in which cases \textit{all} solutions starting
\textit{sufficiently close} to them shall \textit{return} to them),
\textit{unstable equilibria }(in which cases \textit{all} solutions starting
\textit{close} to them shall instead tend to move \textit{away} from them); or
\textit{neutral equilibria }(also called\textit{ indifferent equilibria}), in
which case solutions starting \textit{close} to them turn around them on a
\textit{closed} trajectory, giving rise to a \textit{periodic}, in fact
\textit{isochronous}, $t$-evolution. Of course $\boldsymbol{\bar{x}}^{(+)  }$ and $\boldsymbol{\bar{x}}^{(-)  }$ have the same nature, meaning that
they are either \textit{both real} or \textit{both complex} and, if
\textit{real}, they are \textit{equilibria} of the \textit{same} kind. In
fact, the \textit{linear} system of ODEs (\ref{ydot}) has only $1$ equilibrium
point $\boldsymbol{\bar{y}}\equiv\left(  u_{1},u_{2}\right)  $, and the 2 equilibria
$\boldsymbol{\bar{x}}^{(+)  }$ and $\boldsymbol{\bar{x}}^{(-)  }$ are
the mapping of $\boldsymbol{\bar{y}}$ via (\ref{Solx1x2(t)}). Hence $\boldsymbol{\bar{x}}^{(+)  }$ and $\boldsymbol{\bar{x}}^{(-)  }$ are
\textit{stable}, \textit{unstable} or \textit{neutral} equilibria of the
\textit{nonlinear} system of ODEs (\ref{ODEs}) if $\boldsymbol{\bar{y}}$ is a
\textit{stable}, \textit{unstable} or \textit{neutral} equilibrium of the
\textit{linear} system of ODEs (\ref{ydot}).

We provide below examples of all these types of solutions, being included within
our class of \textit{explicit} solutions of the subclass of the system of ODEs
(\ref{ODEs}) on which we focus. We will classify these solutions according to
the behavior of the solutions of the corresponding ODEs (\ref{ydot}).

We stress that, since the solutions  of the system of ODEs (\ref{ydot}) satisfied by the $y_n(t)$ variables is \textit{linear}, when this system is \textit{isochonous}, \textit{all} its solutions are described by \textit{closed}\textit{ trajectories} in the $y_1-y_2$ plane, and the  \textit{ neutral equilibra} are identified by the \textit{points}  $\boldsymbol{\bar{y}}\equiv\left(  u_{1},u_{2}\right)$, which shall be located \textit{inside} these \textit{trajectories}. The situation for the system of ODEs (\ref{ODEs}) shall be analogous, hence to some \textit{isochronous} solutions of the system (\ref{ydot}) there shall correspond \textit{isochronous} solutions of the system of ODEs (\ref{ODEs}). But there is an important difference. While all the solutions of the \textit{linear} system of ODEs (\ref{ydot}) satisfied by the variables $y_n(t)$ are of course \textit{nonsingular} and may be in some cases \textit{all periodic}, indeed \textit{isochronous}, the solutions of the \textit{nonlinear} system of ODEs (\ref{ODEs}) satisfied by the variables $x_n(t)$ might become \textit{singular} by blowing up at some finite time or by becoming no more well-determined. The \textit{first} possibility is possible for the system of ODEs (\ref{ODEs}) in the \textit{general} case, but does \textit{not} happen for the subclass of \textit{explicitly solvable} systems that we identify and on which we focus in this paper. On the other hand, the \textit{second} phenomenon may happen even when the corresponding system of ODEs (\ref{ydot}) satisfied by the variables $y_n(t)$ is \textit{periodic}, indeed \textit{isochronous}, which might suggest that the corresponding system (\ref{ODEs}) satisfied by the variables $x_n(t)$ shares the same properties.  This indeed does happen in some cases, and we will exhibit specific examples when this happens; but there are instead cases when the system of ODEs (\ref{ODEs}), while corresponding to the system of ODEs (\ref{ydot}), which is \textit{isochronous}, does not itself share this property, because its solutions become \textit{undetermined} after some finite value $t=t_S$ of the independent variable $t$ (assuming that the initial values problem is assigned at the time $t=0$), because the right-hand-side of the ODEs (\ref{ODEs}) becomes, for these solutions, \textit{singular} at $t=t_S$, due to the vanishing of the denominator $P_1(x_1,x_2)$.

\section{Examples}\label{Section examples}
In this \textbf{Section} we report and display examples of systems of ODEs (\ref{ODEs}) which are \textit{explicitly solvable} with the technique described in \textbf{Section 2}.

\subsection{Example 1}\label{Section Example 1}

An interesting subcase is that corresponding to the choice $h_{0}=1$, and
$h_{1}=h_{2}=0$ implying that the right-hand sides of the 2 ODEs (1) are \textit{just polynomials}\textit{ of third degree} (no denominators). So this is the case in which the right-hand sides of
the system of ODEs (\ref{ODEs}) is \textit{polynomial} (at most of the third degree).
In this case, the \textit{constraints} among the $f_{k}$ and $g_{k}$ \textit{coefficients} are
given by eqs. (\ref{12Constrains 2}).

\textbf{Remark 3.1-1.} The \textit{solvability} of the \textit{subcase} of this system with the restriction that the \textit{cubic} \textit{polynomials} on the right-hand-side of the ODEs (\ref{ODEs}) be \textit{homogeneous}---entailing a reduction of the number of their coefficients from $10+10=20$ to $4+4=8$---was previously investigated in Ref. \cite{calogero payandeh} via a variant of the technique of Ref. \cite{G1960}. This special case is \textit{not} included in the treatment of the present paper, since it involves the \textit{a priori} requirement that only the 8 \textit{coefficients} $f_s$ and $g_s$ with $s=9,8,7,6$ do not vanish. $\blacksquare$

By the inversion of eqs. (\ref{f1f2g}) described in \textbf{Subsection \ref{case h10= h2=0} } for
$h_{1}=h_{2}=0$, we get%
\begin{equation}
\bar{\eta}=\left[  -3f_{1}\left(  g_{9}\right)  ^{1/3}/\left(  g_{6}\right)
^{1/3}-\left(f_{3}\right)^2\left(  g_{9}\right)  ^{2/3}/\left(  g_{6}\right)  ^{5/3}%
+f_{3}f_{5}/g_{6}+3f_{2}\right]  /4
\end{equation}

\begin{align}
&  \tilde{\eta}^{2}=\left\{  \left[  \left(  g_{6}\right)  ^{4}g_{9}\right]
^{1/3}\left[  -2f_{2}f_{3}f_{5}g_{6}+\left[  \left(  f_{2}\right)
^{2}-16f_{0}f_{5}\right]  \left(  g_{6}\right)  ^{2}+f_{3}\left(
f_{5}\right)  ^{2}\right]  \right. \nonumber\\
&  -2\left[  g_{6}\left(  g_{9}\right)  ^{1/3}\right]  ^{2}\left[
f_{1}\left(  f_{2}g_{6}+7f_{3}f_{5}\right)  -8f_{3}g_{0}g_{6}\right]
+6f_{1}f_{3}\left(  g_{6}\right)  ^{4/3}\left(  g_{9}\right)  ^{4/3}%
\nonumber\\
&  +f_{3}\left(  g_{9}\right)  ^{5/3}+\left(  g_{6}\right)  ^{2/3}g_{9}\left[
2f_{2}f_{3}g_{6}+\left[  \left(  f_{1}\right)  ^{2}+16f_{0}f_{3}\right]
\left(  g_{6}\right)  ^{2}-2f_{5}f_{3}\right] \nonumber\\
&  \left.  +8f_{5}\left(  g_{6}\right)  ^{8/3}\left(  f_{1}f_{5}-2g_{0}%
g_{6}\right)  \right\}  /\left[  16\left(  g_{6}\right)  ^{10/3}\left(
g_{9}\right)  ^{1/3}\right]  \,,
\end{align}
where $\bar{\eta}$ and $\tilde{\eta}$ are defined in eqs. (\ref{eta bar}) and (\ref{eta tilde}) respectively. Hence, by choosing the \textit{coefficients} $f_{k}$ and $g_{k}$ properly, one can easily
identify a \textit{subclass} of systems (\ref{ODEs}), with \textit{third-degree polynomial right-hand sides},
which is \textit{isochronous}.

For instance, setting%
\begin{equation}
f_{2}=-\left[  -3f_{1}\left(  g_{9}/g_{6}\right)  ^{1/3}-\left(f_{3}\right)^2\left(
g_{9}\right)  ^{2/3}/\left(  g_{6}\right)  ^{5/3}+f_{3}f_{5}/g_{6}\right]
/3\,,
\end{equation}
and choosing $f_{3}=1$, $f_{5}=0$, $f_{0}=-1$, $f_{1}=-1$, $g_{0}=1$,
$g_{9}=1$, $g_{6}=1$, $h_{0}=1$, $h_{1}=0$, and $h_{2}=0$, we get $\bar{\eta
}=0$, $\tilde{\eta}=\textbf{i}~\sqrt{7/18}$, and $\omega=\sqrt{7/18}$, where $\omega$ is defined in eq. (\ref{omega}). Then

\begin{subequations}
\begin{equation}
P_{3}^{(1)}=-\left(  x_{1}\right)  ^{3}-3\left(  x_{1}\right)^{2}x_{2}-3x_{1}\left(  x_{2}\right)^{2}-\left(
x_{2}\right)^{3}
+x_{1}x_{2}+\left(  x_{2}\right)  ^{2}-(2/3){x_{1}}-x_{2}-1\,,
\end{equation}

\begin{equation}
P_{3}^{(2)}=\left(  x_{1}\right)^{3}+3\left(  x_{1}\right)^{2}x_{2}+3x_{1}\left(x_{2}\right)^{2}+\left(  x_{2}\right)^{3}
+[\left(  x_{1}\right)^{2}-\left(x_{2}\right)^{2}]/2+(5/6)x_{1}+(2/3)x_{2}+1\,.
\end{equation}

\end{subequations}
As noted in the previous \textbf{Section \ref{Section constraints}}, the $4$ \textit{parameters} $\beta_{2}$, $\gamma$,
$\eta_{22}$, and $\mu$ are not fixed by the assignment of the $f_{k}$, $g_{k}$
and $h_{\ell}$ \textit{coefficient}s, indeed we can choose them \textit{arbitrarily}, e. g., setting
$\beta_{2}=1$, $\gamma=1$, $\eta_{22}=2$, and $\mu=1$. With this choice, the solution (\ref{Solyn(t)}) of the linear system (\ref{ydot}) is
\begin{align}
&  y_{1}(t)=\left[  -\sqrt{42}\left(  392y_{1}(0)+79y_{2}(0)-513\right)
\sin\left[\omega t\right]  \right. \nonumber\\
&  \left.  +28\left[  49y_{1}(0)+5\right]  \cos\left[\omega t\right]  -140\right]  /1372\,,
\end{align}
\begin{align}
&  y_{2}(t)=\frac{2}{3}\sqrt{\frac{2}{7}}\left[  49y_{1}(0)+9y_{2}%
(0)-58\right]  \sin\left[\omega t\right]  +\left(  y_{2}(0)-7\right)  \cos\left[\omega t\right]  +7\,,
\end{align}
where the \textit{initial} values
$y_{n}\left(  0\right)  $ are expressed in terms of the \textit{initial} values
$x_{n}\left(  0\right)$ by the formulas (\ref{y120}) as follows:
\begin{align}
&  y_{1}(0)=\left\{  -3\left[  x_{1}(0)\right]  ^{2}%
-6x_{2}(0)x_{1}(0)+11x_{1}(0)-3\left[  x_{2}(0)\right]  ^{2}+14x_{2}%
(0)+14\right\}/14  ,\nonumber\\
&  y_{2}(0)=\left[  -14x_{1}(0)-14x_{2}(0)+3\right]/3  \,.
\end{align}

It turns out that, when $h_{0}=1$, $h_{1}=h_{2}=0$, eq. (\ref{C2}) gives
$C_{2}=0$. Indeed, the $x_{n}\left(  t\right)  $ variables are expressed in
terms of the $y_{n}\left(  t\right)  $ variables as in eqs.
(\ref{Solx1x2(t) 2}). This is in agreement with the fact that, when
$P_{1}(x_{1},x_{2})\equiv1$, the solutions of the system of ODEs (\ref{ODEs}) on which we focus in this paper
are \textit{not} singular, and of course do not  become \textit{indeterminate} at \textit{finite} values of the independent variable $t$.  Indeed,  the solution of (\ref{ODEs}) will be%
	\label{solution example x 6}
\begin{subequations}
	\label{solution example 6}
\begin{align}
&  x_{1}(t)=\left\{  -36\sqrt{14}\left(  y_{2}\left(  0\right)  -7\right)
\left(  49y_{1}\left(  0\right)  +9y_{2}\left(  0\right)  -58\right)
\sin\left[2 \omega t\right]  \right. \nonumber\\
&  \left.  -4\sqrt{14}\left(  6664y_{1}\left(  0\right)  +1077y_{2}\left(
0\right)  -6859\right)  \sin\left[\omega t\right]
\right. \nonumber\\
&  \left.  -112\left(  343y_{1}\left(  0\right)  +114y_{2}\left(  0\right)
-763\right)  \cos\left[\omega t\right]  \right.
\nonumber\\
&  \left.  -3\left(  19208y_{1}\left(  0\right)  +784\left(  9y_{2}\left(
0\right)  -58\right)  y_{1}\left(  0\right)  +9y_{2}\left(  0\right)  \left(
79y_{2}\left(  0\right)  -1026\right)  +36887\right)  \right. \nonumber\\
&  
 +3 \cos\left[2 \omega t\right] \left[   19208y_{1}\left(  0\right)  +784\left(  9y_{2}\left(
0\right)  -58\right)  y_{1}\left(  0\right)  
+45y_{2}\left(  0\right)  \left(
13y_{2}\left(  0\right)  -166\right) \right.
\nonumber\\
&  \left. \left.
 +23825\right]    \right\}  /8232,\nonumber\\
\end{align}
\begin{align}
&  x_{2}(t)=\left\{  36\sqrt{14}\left(  y_{2}\left(  0\right)  -7\right)
\left(  49y_{1}\left(  0\right)  +9y_{2}\left(  0\right)  -58\right)
\sin\left[2 \omega t\right]  +\right. \nonumber\\
&    4\sqrt{14}\left(  4606y_{1}\left(  0\right)  +699y_{2}\left(
0\right)  -4423\right)  \sin\left[\omega t\right]\nonumber\\
&  
+28\left(  1372y_{1}\left(  0\right)  +393y_{2}\left(  0\right)  -2611\right)
\cos\left[\omega t\right]   \nonumber\\
&   +3\left[  19208y_{1}\left(  0\right)  +784\left(  9y_{2}\left(
0\right)  -58\right)  y_{1}\left(  0\right)  +9y_{2}\left(  0\right)  \left(
79y_{2}\left(  0\right)  -1026\right)  +33359\right]   \nonumber\\
&   -3 \cos\left[2 \omega t\right] \left[  19208y_{1}\left(  0\right)  +784\left(  9y_{2}\left(
0\right)  -58\right)  y_{1}\left(  0\right)  +45y_{2}\left(  0\right)  \left(
13y_{2}\left(  0\right)  -166\right) \right.  \nonumber\\
&\left. \left.+23825\right]   \right\}  /8232\,.
\end{align}
\end{subequations}

\begin{figure}
	\begin{center}
		\includegraphics[height=7cm]{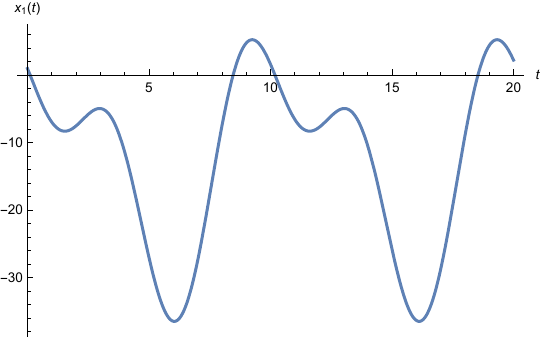}\\
		\vskip.5cm
		\includegraphics[height=7cm]{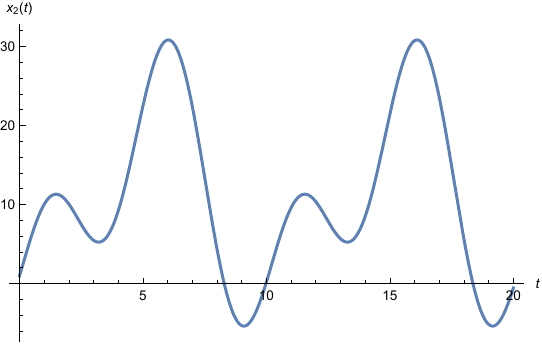}
		\caption{We plot the functions $x_{1}(t)$ and $x_{2}(t)$ of eq. (\ref{solution example 6}) for the initial value $x_1(0)=x_2(0)=1$. 	}		\label{Ex6fig1}
	\end{center}
\end{figure}

\begin{figure}
	\begin{center}
		\includegraphics[height=11cm]{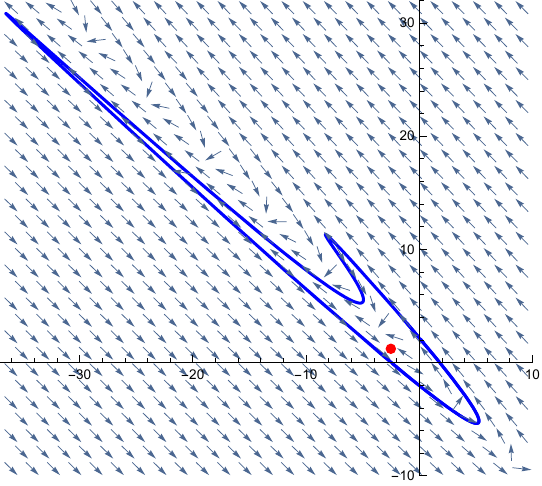}
		\caption{We plot in red  the trajectory  in the $x_{1}-x_{2}$ plane corresponding to the solution (\ref{solution example 6}) for the initial value $x_1(0)=x_2(0)=1$, together with  the equilibrium  point in red, and
			the velocity field $\boldsymbol{v}= \left(  \dot x_{1}, \dot x_{2}\right)$ indicated by the arrows in the background.
		}		\label{Ex6fig2}
	\end{center}
\end{figure}

An example of the solution (\ref{solution example 6})  for the initial values $x_{1}(0)=x_{2}(0)=1$ is plotted in Fig. \ref{Ex6fig1}, while the corresponding trajectory in the $x_{1}-x_{2}$ plane is plotted in Fig. \ref{Ex6fig2}, together with the equilibrium point and the velocity field $\boldsymbol{v}=\left(\dot{x}_1,\dot{x}_2\right)$.

\bigskip

\subsection{Example 2}

\label{Example 1}

	In this \textbf{Subsection 3.2} we consider the situation in which the \textit{linear} system of ODEs (\ref{ydot}) satisfied by the variables $y_n(t)$ is \textit{isochronous}, while the system of ODEs (\ref{ODEs}) satisfied by the variables $x_n(t)$ features solutions which are \textit{isochonous}, as well as solutions which become \textit{undetermined} after some \textit{finite} value $t=t_S$ of the independent variable $t$, because the polynomial $P_1(x_1,x_2)$ in the denominators of the system of ODEs (\ref{ODEs}) vanishes at $t=t_S$, i. e., $P_1(x_1(t_S),x_2(t_S))=0$.

 As
suggested by the results reported in \textbf{Subsection \ref{section  isochronous case}}, we set
\begin{equation}
\eta_{11}=-\eta_{22}=a~,~~\eta_{12}=b~,~~\eta_{21}=-\left(  a^{2}+\omega
^{2}\right)  /b~, \label{eta parameters periodic}%
\end{equation}
with $a,b$ and $\omega$ \textit{real} parameters: this ensures that the
\textit{parameters} $\phi^{(\pm)}$ defined by eq. (\ref{phi pm}) are \textit{imaginary}%
, i.e., $\phi^{(\pm)}=\pm \textbf{i}\,\omega$, hence the $y_{n}(t)$ solutions of
(\ref{ydot}) are \textit{periodic}, their trajectories in the $y_{1}%
-y_{2}$ Cartesian plane being of course \textit{ellipses}. However, these
conditions are \textit{not sufficient} to imply that the corresponding
$x_{n}(t)$ in (\ref{Solx1x2(t)}) are \textit{periodic}, as their trajectories
can intersect the line $P_{1}(x_{1},x_{2})=0$ where the system of ODEs
(\ref{ODEs}) becomes\textit{\ singular }(because the \textit{denominator} in
the right-hand sides of these ODEs \textit{vanishes}).

Hereafter we set $h_{0}=h_{2}=0$, so that $P_{1}(x_{1},x_{2})=h_{1}x_{1}$, then the \textit{singularity line} will coincide with the $x_{1}=0$
\textit{axis}. This is convenient to present graphically the examples.
Moreover, one can always reduce to this case with an appropriate shift and
rotation of the coordinate in the $x_{1}-x_{2}$ Cartesian plane (except in the
exceptional case when \ $h_{1}=h_{2}=0$, discussed in \textbf{Subsection 3.1}).

Imposing these conditions, and solving eqs. (\ref{h0}), (\ref{h1}%
), (\ref{h2}) for the \textit{parameters} $\lambda_{1},\lambda_{2},\beta_{1}$, we get
\begin{subequations}
\label{lambda h conditions}%
\begin{equation}
\lambda_{1}=-\alpha_{1}h_{1}/\left[  \left(  \alpha_{1}\right)  ^{2}%
-4\alpha_{0}\alpha_{2}\right]  ~,
\end{equation}

\begin{equation}
\lambda_{2}=-2\alpha_{2}h_{1}/\left[  \left(
\alpha_{1}\right)  ^{2}-4\alpha_{0}\alpha_{2}\right]  ~,
\end{equation}

\begin{equation}
\beta_{1}=\alpha_{1}\beta_{2}/\left(  2\alpha
_{2}\right)  ~\,,~
\end{equation}
which are equivalent to eqs. (\ref{inversion lambda beta1}). We then set%

\end{subequations}
\begin{equation}
a=-1~,~b=2~,~\omega=2~,
\end{equation}
so that eqs. (\ref{eta parameters periodic}) give
\begin{subequations}
\label{parameters example 1}%
\begin{equation}
\eta_{11}=-\eta_{22}=-1~,~~\eta_{12}=2~,~~\eta_{21}=-5/2~.
\end{equation}
Moreover, we set
\begin{align}
\xi_{1}  &  =0.2~,~~\xi_{2}=-0.3~,~~\beta
_{2}=-0.3~,~~\gamma=-0.1~,~~\mu=0.5~,\nonumber\\
\alpha_{0}  &  =-0.1~,~~\alpha_{1}=0~,~~\alpha_{2}=0.3~,~~h_{1}=1,~
\end{align}
so that the eqs. (\ref{lambda h conditions}) give%
\begin{equation}
\lambda_{1}=0~,~~\lambda_{2}=-5~,~~\beta
_{1}=0\mathbf{~.} 
\end{equation}
Then, from the eqs. (\ref{f1f2g}), we get
\end{subequations}
\begin{subequations}
\label{Example 1 polynomials}%
\begin{align}
&  P_{3}^{(1)}(x_{1},x_{2})=\left[-30\left(  x_{1}\right)  ^{2}x_{2}+  90\left(  x_{2}\right)  ^{3}
-85\left(  x_{1}\right)
^{2}+765\left(  x_{2}\right)  ^{2}\right. \nonumber\\
&  \left.  +9391x_{2}-1273\right]  /200~,
\end{align}%
\begin{equation}
P_{3}^{(2)}(x_{1},x_{2})=\left\{  5\left[
-\left(  x_{1}\right)  ^{2}+3\left(  x_{2}\right)  ^{2}+17x_{2}\right]
-9\right\}  x_{1}/100~,
\end{equation}%
\begin{equation}
P_{1}(x_{1},x_{2})=x_{1}~;
\end{equation}
so in this example $\dot{x}_{2}$ is \textit{never} singular, because the right-hand-side of eq. (\ref{xx12dot}) with $n=2$ becomes just a polynomial of \textit{second} degree.

From eqs. (\ref{u1}) and (\ref{u2}) we get $u_{1}=-0.2$ and $u_{2}=-0.2$, and the points
\end{subequations}
\begin{subequations}
\label{Example1 equilibrium points}%
\begin{equation}
\boldsymbol{\bar{x}^{(+)}}=\left(  X_{1}^{(+)}\left(  u_{1},u_{2}\right)  ,X_{2}^{\left(
+\right)  }\left(  u_{1},u_{2}\right)  \right)  =\left(  \sqrt{1597}%
/50,7/50\right)  \,,
\end{equation}
\begin{equation}
\boldsymbol{\bar{x}^{(-)}}=\left(  X_{1}^{(-)}\left(
u_{1},u_{2}\right)  ,X_{2}^{\left(  -\right)  }\left(  u_{1},u_{2}\right)
\right)  =\left(  -\sqrt{1597}/50,7/50\right)
\end{equation}
in the $x_1-x_2$ Cartesian plane identify 2 \textit{equilibrium} solutions of the system (\ref{ODEs}). Since the point $\boldsymbol{\bar{y}}=(u_{1},u_{2})$ of the $y_1-y_2$ Cartesian plane identifies a
\textit{neutral equilibrium} solution of the system (\ref{ydot}),  the corresponding points $\boldsymbol{\bar{x}^{(+)}}$ and $\boldsymbol{\bar{x}^{(-)}}$ will be themselves \textit{neutral equilibrium} solutions.

For the choice of \textit{parameters} in eqs. (\ref{parameters example 1}), the
functions $X_{n}^{(  \pm)  }$ defined\ by the
eqs. (\ref{Solx1x2(t)}) are
\end{subequations}
\begin{subequations}
\label{solution x example 1}
\begin{align}\label{solution x example 1 x1}
&  X_{1}^{(\pm)}\left(y_1(t),y_2(t)\right)  =\mp\left[  -48\left[  25y_{1}\left(
0\right)  -10y_{2}\left(  0\right)  +3\right]  \left(  5y_{2}\left(  0\right)
+1\right)  \sin(4t)\right. \nonumber\\
&  \left. + 32\left[  5575y_{1}\left(  0\right)  -12230y_{2}\left(  0\right)
-1331\right]  \sin(2t)\right. \nonumber\\
&  \left.  -6\left[  25y_{1}\left(  0\right)  -30y_{2}\left(  0\right)
-1\right]  \left[  25y_{1}\left(  0\right)  +10y_{2}\left(  0\right)
+7\right]  \cos(4t)\right. \nonumber\\
&  \left.  -128\left[  3125y_{1}\left(  0\right)  -135y_{2}\left(  0\right)
+598\right]  \cos(2t)+3750\left[  y_{1}\left(  0\right)  \right]
^{2}-3000y_{2}\left(  0\right)  y_{1}\left(  0\right)  \right. \nonumber\\
&  \left.  +900y_{1}\left(  0\right)  +3000y_{2}\left(  0\right)
^{2}+600y_{2}\left(  0\right)  +25702\right]  ^{1/2}\,~/200,
\end{align}%
\begin{align}
&  X_{2}^{(\pm)}\left(y_1(t),y_2(t)\right)  =\left\{
{}\right.  25y_{1}\left(  0\right)  -10y_{2}\left(  0\right)  +3\sin
(2t)\nonumber\\
&  -4\left[  5y_{2}\left(  0\right)  +1\right]  \cos(2t)+14\left.  {}\right\}
\,/100~,
\end{align}
where the \textit{initial values} $y_{n}(0)$ are given by the expressions
(\ref{y120}) in terms of the \textit{initial values} $x_{n}(0)$:
\end{subequations}
\begin{subequations}
\label{y120 Example 1}%
\begin{equation}
y_{1}\left(  0\right)  =\left\{  -\left[  x_{1}(0)\right]  ^{2}+3\left[
x_{2}(0)\right]  ^{2}-3\left[  x_{2}(0)\right]  ^{2}-1\right\}  /10~,
\end{equation}%
\begin{equation}
y_{2}\left(  0\right)  =-5x_{2}(0)/2~.
\end{equation}
For \textit{initial values sufficiently close} to the \textit{equilibrium
point}, we get\textit{ periodic} solutions. For instance, for $x_{1}(0)=-0.99$
and $x_{2}(0)=0.14$, we get $y_{1}(0)=-0.23413$ and $y_{2}(0)=-0.2$. Since the
\textit{initial value} $x_{1}(0)$ is \textit{negative}, we have to choose the
\textit{negative} sign in (\ref{solution x example 1 x1}), so that the solution
is
\end{subequations}
\begin{subequations}
\label{Example 1 x(t) periodic}%
\begin{align}
x_{1}(t)  &  =-\left[  40.8902-9.74207\sin(2t)+21.8432\cos(2t)\right.
\nonumber\\
&  \left.  -6.98914\cdot10^{-3}\cos(4t)\right]  ^{1/2}\,/8 ~,
\end{align}%
\begin{equation}
x_{2}(t)=0.14\,-8.5325\cdot10^{-3}\sin(2t)\,.
\end{equation}
The functions $x_{1}(t)$ and $x_{2}(t)$ of eq. (\ref{Example 1 x(t) periodic})
are plotted in fig. \ref{Ex1fig1}.

\begin{figure}
	\begin{center}
		\includegraphics[height=7cm]{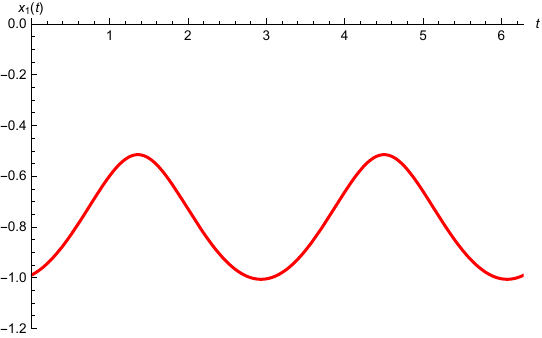}\\
		\includegraphics[height=7cm]{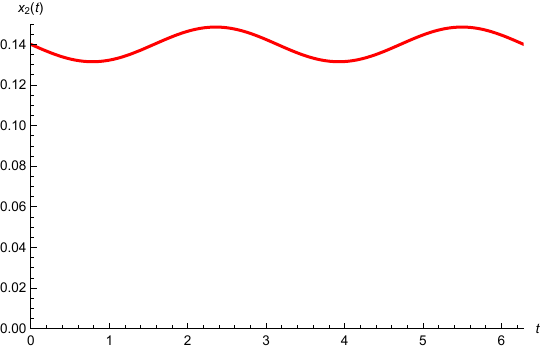}
		\caption{We plot the functions $x_{1}(t)$ and $x_{2}(t)$ of eqs. (\ref{Example 1 x(t) periodic}).}		\label{Ex1fig1}
	\end{center}
\end{figure}

\begin{figure}
	\begin{center}
		\includegraphics[height=7cm]{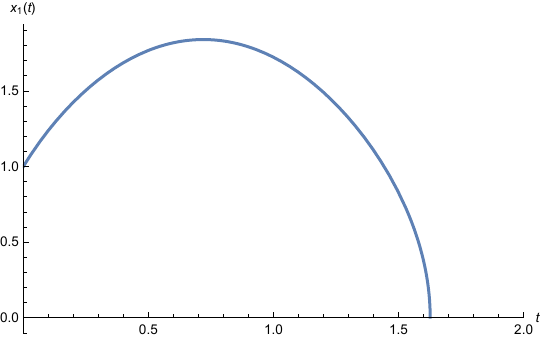}\\
		\includegraphics[height=7cm]{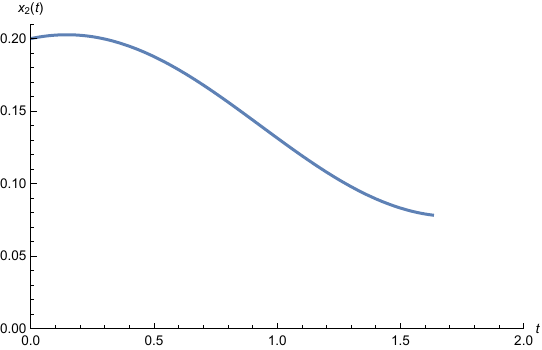}
		\caption{We plot the functions $x_{1}(t)$ and $x_{2}(t)$ of eqs.
			(\ref{Example 1 x(t) non periodic}). The corresponding solution of the system of 2 ODEs (\ref{ODEs}) becomes
			\textit{undetermined} when $x_{1}(t)=0$.}		\label{Ex1fig2}
	\end{center}
\end{figure}

\begin{figure}
	\begin{center}
		\includegraphics[height=12cm]{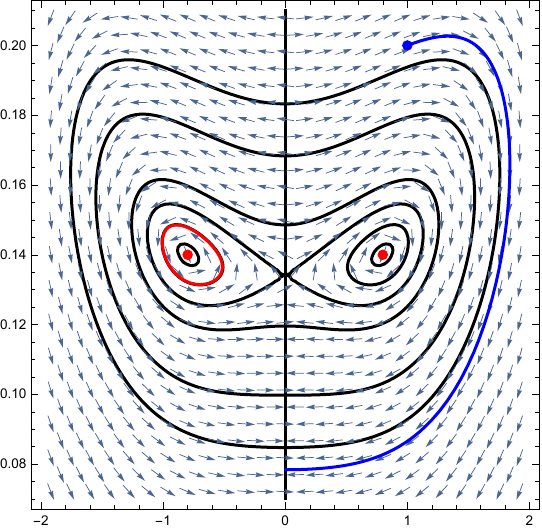}
		\caption{We plot the
			trajectories of the 2 solutions (\ref{Example 1 x(t) periodic}) and
			(\ref{Example 1 x(t) non periodic})  in the $x_{1}-x_{2}$ plane,
			respectively in \textit{red} and \textit{blue}. In this figure we plot in
			\textit{red} the $2$ equilibrium points $\boldsymbol{\bar{x}^{(+)}}$ and $\boldsymbol{\bar{x}^{(-)}}$
			 in eqs. (\ref{Example1 equilibrium points}), and, in \textit{black},
			11 other \textit{periodic} and \textit{non-periodic} solutions,
			together with the velocity field $\boldsymbol{v}=\left(  \dot{x}_{1},\dot{x}_{2}\right)  $, and the \textit{singularity} line $P_{1}(x_{1},x_{2})=0$. From
			these plots it is evident that some of the trajectories, those \textit{close} to
			the \textit{equilibrium} points, are \textit{periodic}, while other curves
			start and finish on the \textit{singularity} line at a \textit{finite} time,
			when they \textit{cease} to be \textit{genuine} solutions of the system of ODEs (\ref{ODEs}).}		\label{Ex1fig3}
	\end{center}
\end{figure}

Instead, for \textit{initial values sufficiently far} from the
\textit{equilibrium point}, we get solutions which become \textit{undetermined} at some \textit{finite} time $t=t_S$, when they touch the line $P_1(x_1,x_2)=0$. For instance, the \textit{initial values} 
$x_{1}(0)=1$ and $x_{2}(0)=0.2$ give $y_{1}(0)=-0.248$ and $y_{2}(0)=-0.2$,
and the corresponding solution is%
\end{subequations}
\begin{subequations}
\label{Example 1 x(t) non periodic}%
\begin{align}
x_{1}(t)  &  =\left[  2.72112\sin(2t)+0.00324\sin(4t)+0.3504\cos
(2t)+0.004914\cos(4t)\right. \nonumber\\
&  \left.  +0.644686\right]  ^{1/2}\ ~,
\end{align}
\begin{equation}
x_{2}(t)=\left[  (9\sin(2t)+30\cos
(2t)+70\right]  /\,500 ~,
\end{equation}
as we now have to choose the \textit{positive} sign in
(\ref{solution x example 1}), since the initial value $x_{1}(0)$ is \textit{positive}.
The functions $x_{1}(t)$ and $x_{2}(t)$ in eq.
(\ref{Example 1 x(t) non periodic}) are plotted in Fig. \ref{Ex1fig2}. They  cease to be a proper solution of the system of ODEs (\ref{ODEs})  when $x_{1}(t)=0$, that is, when the corresponding
trajectory hits the \textit{singularity} line $P_{1}(x_{1},x_{2})=0$.

In Fig. \ref{Ex1fig3}, the trajectories of the 2 solutions
(\ref{Example 1 x(t) periodic}) and (\ref{Example 1 x(t) non periodic}) are
plotted in the $x_{1}-x_{2}$ plane, respectively in \textit{red} and
\textit{blue}. In the same figure we plot in \textit{red} the $2$ \textit{equilibrium}
points $\bar{x}_{1}$ and $\bar{x}_{2}$ defined by eqs.
(\ref{Example1 equilibrium points}), and in \textit{black}, 11 other
\textit{periodic} and \textit{non-periodic} solutions, together with the
velocity field $\boldsymbol{v}=\left(  \dot{x}_{1},\dot{x}_{2}\right)$, and the
\textit{singularity} line $P_{1}(x_{1},x_{2})=0$. From this plot it is evident
that some of the trajectories, those \textit{close} to the
\textit{equilibrium} points, are \textit{periodic}, while other curves start
and finish on the \textit{singularity} line at a \textit{finite} time, when
they \textit{cease} to be \textit{genuine} solutions of the system of ODEs (\ref{ODEs}).

\bigskip

\subsection{Example 3}

In this third example, we still set the $\eta_{nm}$ parameters as in eqs.
(\ref{eta parameters periodic}), so that the solutions of the \textit{linear}
system of ODEs (\ref{ydot}) are \textit{periodic} indeed \textit{isochronous}. Moreover, we also take
$\lambda_{1}$, $\lambda_{2}$, and $\beta_{1}$ as in eqs.
(\ref{lambda h conditions}), so that again $h_{0}=h_{2}=0$.

We have seen in the previous example that the fact that the $y_{n}(t)$ are
\textit{periodic} does not imply necessarily that the solutions of the ODEs
(\ref{ODEs}) $x_{n}(t)=X_{n}^{\left(  \pm\right)  }[y_{1}(t),y_{2}(t)]$ are
also \textit{periodic}, as they may cease to be solutions at some
\textit{finite} time $t_{S}$, when $P_{1}(x_{1},x_{2})=x_{1}=0$.~In that case, depending on the initial data, while all solutions $x_{n}(t)$ might have appeared to be periodic (indeed \textit{isochronous}), this was not the case since some might instead become \textit{undetermined} at some finite positive value of the independent value $t=t_s$.

 As we will
show below, in the present example, there are no \textit{equilibria} nor
\textit{periodic} solutions: \textit{all} the trajectories start and finish
on the \textit{singularity} line $x_{1}=0$; hence for \textit{any} initial data the solutions become \textit{undetermined} at some \textit{finite positive} value of the \textit{independent} variable $t=t_s<T$, itself of course depending on the initial data.

We set%

\end{subequations}
\begin{equation}
a=1~,~b=2,~\omega=1,
\end{equation}
so that eqs. (\ref{eta parameters periodic}) give
\begin{subequations}
\label{parameters example 2}%
\begin{equation}
\eta_{11}=-\eta_{22}=1~,\eta_{12}=2~,\eta_{21}=-1~.
\end{equation}
Moreover, we set
\begin{align}
\xi_{1}  &  =2~,\xi_{2}=-3~,\beta_{2}%
=3~,\gamma=1~,\mu=1~,\nonumber\\
\alpha_{0}  &  =1~,\alpha_{1}=0~,\alpha_{2}=1,~~h_{1}=1,
\end{align}
so that eqs. (\ref{lambda h conditions}) give%
\begin{equation}
\lambda_{1}=0~,\lambda_{2}=1/2~,\beta_{1}=0~,
\end{equation}
which give
\end{subequations}
\begin{subequations}
\label{Example 2 polynomials}%
\begin{align}
P_{3}^{(1)}(x_{1},x_{2})  &  =52\left(  x_{2}\right)  ^{3}+2\left(
x_{1}\right)  ^{2}x_{2}+\left[\left(
x_{1}\right)  ^{2}+  21\left(  x_{2}\right)  ^{2}+45x_{2}+35\right]  /2~,\\
P_{3}^{(2)}(x_{1},x_{2})  &  =-\left[  2\left(  x_{1}\right)  ^{2}+2\left(
x_{2}\right)  ^{2}+7x_{2}+10\right]  x_{1}~,\\
P_{1}(x_{1},x_{2})  &  =x_{1}~,
\end{align}
so also in this example $\dot{x}_{2}$ is \textit{never} singular, see eqs. (\ref{ODEs}).

From eqs. (\ref{u1}) and (\ref{u2}) we get $u_{1}=-4$ and $u_{2}=1$, and the components  of the points

\end{subequations}
\begin{subequations}
\label{Example2 equilibrium points}%
\begin{equation}
\boldsymbol{\bar{x}^{(+)}}=\left(  X_{1}^{(+)}\left[  u_{1},u_{2}\right]  ,X_{2}^{(+)}\left[
u_{1},u_{2}\right]  \right)  =\left(  \textbf{i}\sqrt{5},0\right)  \,,
\end{equation}

\begin{equation}
\boldsymbol{\bar{x}^{(-)}}=\left(  X_{1}^{(-)}\left[
u_{1},u_{2}\right]  ,X_{2}^{(+)}\left[  u_{1},u_{2}\right]  \right)  =\left(
-\textbf{i}\sqrt{5},0\right)
\end{equation}
are now \textit{imaginary}, hence we conclude that there are \textit{no real equilibrium solutions}
of the system (\ref{ODEs}) in this example.

For the parameters in eq. (\ref{parameters example 2}), the functions defined
by the eqs. (\ref{Solx1x2(t)}) are
\end{subequations}
\begin{subequations}
		\label{solutions XX example 2}%
\begin{align}
	\label{solutions X example 2}%
&  X_{1}^{(\pm)}\left[  y_{1}(t),y_{2}(t)\right]  =\pm\left\{  {}\right.
4\left[  y_{2}\left(  0\right)  -1\right]  \left[  y_{1}\left(  0\right)
+y_{2}\left(  0\right)  +3\right]  \sin(2t)\nonumber\\
&  \left.  +\left[  7y_{1}\left(  0\right)  +8y_{2}\left(  0\right)
+20\right]  \sin(t)+2\left[  y_{1}\left(  0\right)  +4\right]  \left[
y_{1}\left(  0\right)  +2y_{2}\left(  0\right)  +2\right]  \cos(2t)\right.
\nonumber\\
&  \left.  +\left[  y_{1}\left(  0\right)  -6y_{2}\left(  0\right)
+10\right]  \cos(t)-2y_{1}\left(  0\right)  ^{2}-4y_{2}\left(  0\right)
y_{1}\left(  0\right)  \right. \nonumber\\
&  -12y_{1}\left(  0\right)  -4y_{2}\left(  0\right)  ^{2}-8y_{2}\left(
0\right)  -25\left.  {}\right\}  ^{1/2},\\
&  X_{2}^{\left(  \pm\right)  }\left[  y_{1}(t),y_{2}(t)\right]  =2\left[
y_{2}\left(  0\right)  -1\right]  \cos(t)-2\left[  y_{1}\left(  0\right)
+y_{2}\left(  0\right)  +3\right]  \sin(t)\,,
\end{align}
where the initial values $y_{n}(0)$ are given by the expressions (\ref{y120})
in terms of the initial values $x_{n}(0)$:%
\end{subequations}
\begin{equation}
y_{1}\left(  0\right)  =\left[  x_{1}(0)\right]  ^{2}+\left[  x_{2}(0)\right]
^{2}+3x_{2}(0)+1~,~~y_{2}\left(  0\right)  =1+x_{2}(0)/2~.~
\end{equation}

We then see that, in this case, the solutions (\ref{solutions XX example 2}) eventually
always cease to be solutions of their system of ODEs for \textit{any initial
values}, as there is always a $t_{S}$ such that $x_{1}(t_{S})=0$. For
instance, for the $2$ \textit{initial} values $x_{1}(0)=1$ and $x_{2}(0)=1$ we get
$y_{1}(0)=6$ and $y_{2}(0)=3/2$. As the initial value $x_{1}(0)$ is
\textit{positive}, we have to choose the \textit{positive} sign in
(\ref{solutions X example 2}), so that the corresponding solution is%
\begin{subequations}
\label{solutions x(t) example 2}%
\begin{equation}
x_{1}(t)=\sqrt{74 \sin(t)+21 \sin(2 t)+7 \cos(t)+220 \cos(2 t)-226}\, ,
\end{equation}
\begin{equation}
x_{2}(t)=\cos(t)-21\sin(t)\,.
\end{equation}
The functions $x_{1}(t)$ and $x_{2}(t)$ in eq. (\ref{solutions x(t) example 2}%
) are plotted in Fig. \ref{Ex2fig1}.

The corresponding trajectory is plotted in Fig. \ref{Ex2fig2} in \textit{red}
in the $x_{1}-x_{2}$ plane, together with $8$ other trajectories in
\textit{black}, the velocity field $\boldsymbol{v}=\left(  \dot{x}_{1},\dot{x}%
_{2}\right)  $, and the \textit{singularity} line $P_{1}(x_{1},x_{2})=0$. From
this plot it is evident that in this case \textit{all } the solutions of the system of ODEs (\ref{ODEs}) eventually cease to be solutions, because they \textit{all} hit the
\textit{singularity} line $P_{1}(x_{1},x_{2})=0$ at the  \textit{finite} time $t_S$.

\begin{figure}
	\begin{center}
		\includegraphics[height=7cm]{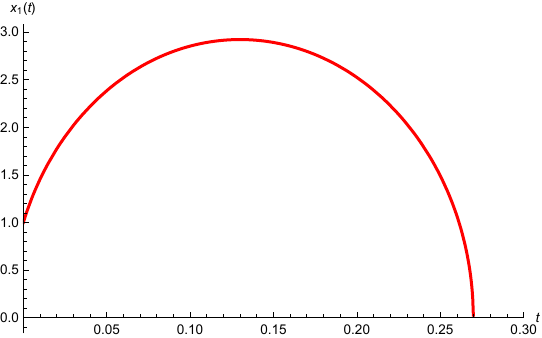}\\
		\includegraphics[height=7cm]{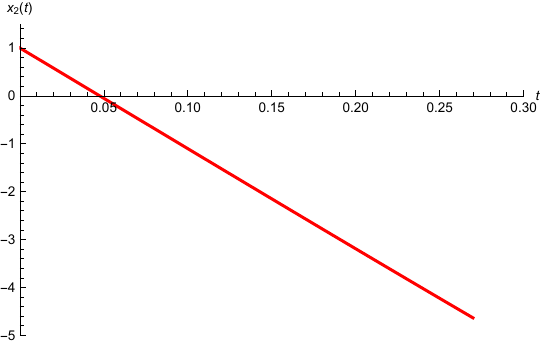}
		\caption{We plot the functions $x_{1}(t)$ and $x_{2}(t)$ of eqs. (\ref{solutions x(t) example 2}). The corresponding solution of the system of 2 ODES (\ref{ODEs})  becomes \textit{undetermined} when $x_1(t)=0$. The function $x_2(t)$ looks like a line, since in the time interval where the solution is plotted, $x_2(t)= 1-21\, t + O(t^2)$. }		\label{Ex2fig1}
	\end{center}
\end{figure}

\begin{figure}
	\begin{center}
		\includegraphics[height=12cm]{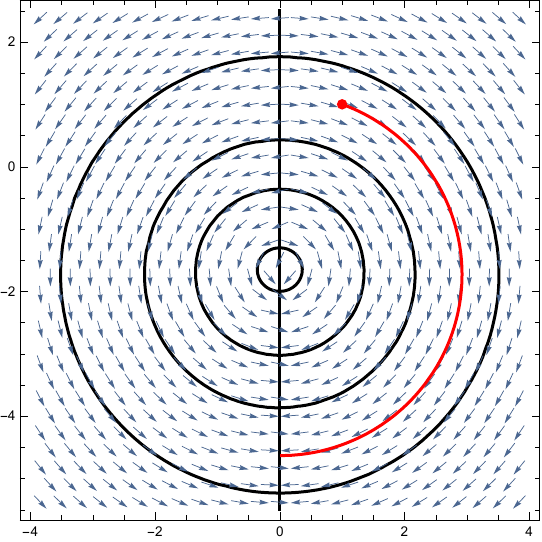}
		\caption{We plot  in \textit{red} the trajectory in the
			$x_{1}-x_{2}$ plane corresponding to the solution (\ref{solutions x(t) example 2}), together with $8$ other trajectories in \textit{black}, the velocity field $\boldsymbol{v}=\left(  \dot{x}_{1},\dot{x}_{2}\right)$, and the \textit{singularity} line $P_{1}(x_{1},x_{2})=0$.
			From this plot it is evident that in this case \textit{all }solutions of the
			system of ODEs (\ref{ODEs}) eventually cease to be solutions because they \textit{all} hit the
			\textit{singularity} line  at the \textit{finite} time $t_S$.
		}		\label{Ex2fig2}
	\end{center}
\end{figure}

\bigskip

\subsection{Example 4}

In this fourth example we consider a case in which the 2 parameters $\phi^{(\pm)}$ defined in eq. (\ref{phi pm}) are \textit{real}, \textit{equal} in
modulus, and have \textit{opposite} signs. We set
\end{subequations}
\begin{equation}
\eta_{11}=-\eta_{22}=a~,~~\eta_{12}=b~,~~\eta_{21}=\left(  \omega^{2}%
-a^{2}\right)  /b~, \label{eta parameters non periodic example 3}%
\end{equation}
so that $\phi^{(\pm)}=\pm\,\omega$, hence the $y_{n}(t)$ solutions in
(\ref{Solyn(t)}) are \textit{hyperboles}. We still take $\lambda_{1}$,
$\lambda_{2}$, and $\beta_{1}$ as in eqs. (\ref{lambda h conditions}), so that
$h_{0}=h_{2}=0$.

We now make the following choice of parameters:%

\begin{equation}
a=1,\, b=1,\,\omega=1,
\end{equation}
so that eqs. (\ref{eta parameters non periodic example 3}) give
\begin{subequations}
\label{parameters example 3}%
\begin{equation}
\eta_{11}=-\eta_{22}=1,\, \eta_{12}=1,\,\eta_{21}=0~.
\end{equation}
Moreover, we set
\begin{align}
\xi_{1}  &  =1,\, \xi_{2}=-1,\,\beta_{2}%
=-1,\,\gamma=1,\,\mu=1,\,\nonumber\\
\alpha_{0}  &  =1,\,\alpha_{1}=0,\,\alpha_{2}=-1,\,h_{1}=1,
\end{align}
so that eqs. (\ref{lambda h conditions}) give%
\begin{equation}
\lambda_{1}=0,\,\lambda_{2}=1/2,\,\beta_{1}=0,
\end{equation}
hence%
\end{subequations}
\begin{subequations}
\label{Example 3 polynomials}%
\begin{align}
P_{3}^{(1)}(x_{1},x_{2})  &  =\left[  2\left(  x_{1}\right)  ^{2}-6\left(
x_{2}\right)  ^{2}-19x_{2}-2\right]  ~/4\,,\\
P_{3}^{(2)}(x_{1},x_{2})  &  =-\left(  x_{2}+4\right)  x_{1}~,\\
P_{1}(x_{1},x_{2})  &  =x_{1}~.
\end{align}
Again, $\dot{x}_{2}$ is \textit{polynomial}, see eqs. (\ref{ODEs}).

In this case we get $u_{1}=0$ and $u_{2}=-1$, see eqs. (\ref{u1}) and (\ref{u2}), and the points
\end{subequations}
\begin{subequations}\label{Example3 equilibrium points}
\begin{align}
\boldsymbol{\bar{x}^{(+)}}  &  =\left(  X_{1}^{(+)}\left(  u_{1},u_{2}\right)  ,X_{2}%
^{(+)}\left(  u_{1},u_{2}\right)  \right)  =\left(  \sqrt{11},-4\right)
\,,\\
\boldsymbol{\bar{x}^{(+)}}   &  =\left(  X_{1}^{(-)}\left(  u_{1},u_{2}\right)  ,X_{2}%
^{(-)}\left(  u_{1},u_{2}\right)  \right)  =\left(  -\sqrt{11},-4\right)  ~,
\end{align}
of the $x_1-x_2$ Cartesian plane identify two \textit{unstable} equilibrium solutions of the system (\ref{ODEs}), since
the point $\boldsymbol{\bar{y}}=(u_{1},u_{2})$ of the $y_1-y_2$ Cartesian plane  is now an \textit{unstable equilibrium solution} of the
system of ODEs (\ref{ydot}).

For the parameters in (\ref{parameters example 3}), we get
\end{subequations}
\begin{subequations}
\label{solution x example 3}%
\begin{align}
	\label{solution xx example 3}%
&  X_{1}^{(\pm)}\left[  y_{1}(t),y_{2}(t)\right]  =\pm\left\{  \left.
{}\right.  \right.  8\exp\left(  -2t\right)  \left[  y_{2}\left(  0\right)
+1\right]  {}^{2}-29\exp\left(  -t\right)  \left[  y_{2}\left(  0\right)
+1\right]  +\nonumber\\
&  +\exp\left(  t\right)  \left[  2y_{1}\left(  0\right)  +y_{2}\left(
0\right)  +1\right]  +22\left.  {}\right\}  ^{1/2}/2\,~,\\
&  X_{2}^{\left(  \pm\right)  }\left[  y_{1}(t),y_{2}(t)\right]  =2\left[
\exp\left(  -t\right)  \left(  y_{1}\left(  0\right)  +1\right)  -2\,\right]
~,
\end{align}
where the initial values $y_{n}(0)$ are given by the expressions (\ref{y120})
in terms of the initial values $x_{n}(0)$, which now read
\end{subequations}
\begin{equation}
y_{1}\left(  0\right)  =\left[  x_{1}(0)\right]  ^{2}-\left[  x_{2}(0)\right]
^{2}-x_{2}(0)+1~,~~~y_{2}\left(  0\right)  =x_{2}(0)/2+1~.
\label{y120 Example 3}%
\end{equation}
For \textit{initial values} $x_{1}(0)=3$ and $x_{2}(0)=2$, we get $y_{1}(0)=4$
and $y_{2}(0)=2$. Since the \textit{initial value} $x_{1}(0)$ is \textit{positive}, we
must choose the \textit{positive} sign in (\ref{solution xx example 3}), so
that the corresponding solution of the ODEs (1) is%

\begin{subequations}
\label{solutions x(t) example 3}%
\begin{equation}
x_{1}(t)=\left\{  49\sinh(t)+36\cosh(2t)-2\left[  (36\sinh(t)+19\right]
\cosh(t)+11\right\}  ^{1/2}\,,
\end{equation}
\begin{equation}
x_{2}(t)=6\exp\left(  -t\right)  -4\,.
\end{equation}
The functions $x_{1}(t)$ and $x_{2}(t)$ of eq. (\ref{solutions x(t) example 3}%
) are plotted in Fig. \ref{Ex3fig1}. The corresponding \textit{trajectory} in
the $x_{1}-x_{2}$\textit{\ }plane is plotted in \textit{red} in Fig.
{\ref{Ex3fig2}}, together with several other\textit{ trajectories} in
\textit{black}, with the $2$\textit{\ unstable equilibrium points} $\boldsymbol{\bar{x}^{(+)}}$ and $\boldsymbol{\bar{x}^{(-)}}$ defined in eqs. (\ref{Example3 equilibrium points}), and the \textit{velocity} field $\boldsymbol{v}=\left(  \dot{x}_{1},\dot{x}%
_{2}\right)  $. From these plots it is also evident that the solution
(\ref{solutions x(t) example 3}) is non-singular, as it never hits the
singularity line $P_{1}(x_{1},x_{2})=0$, and it can be extended indefinitely
in the past and in the future. However, \textit{other trajectories} can meet the
singularity line at a \textit{finite} time.

\begin{figure}
	\begin{center}
		\includegraphics[height=7cm]{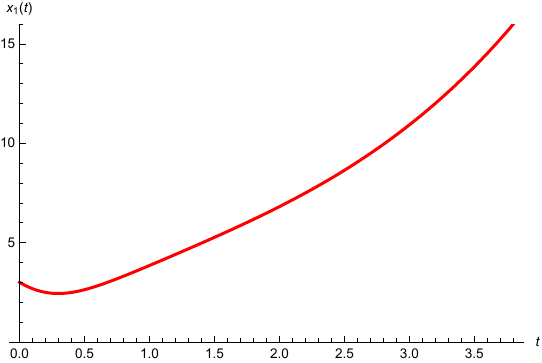}\\
		\includegraphics[height=7cm]{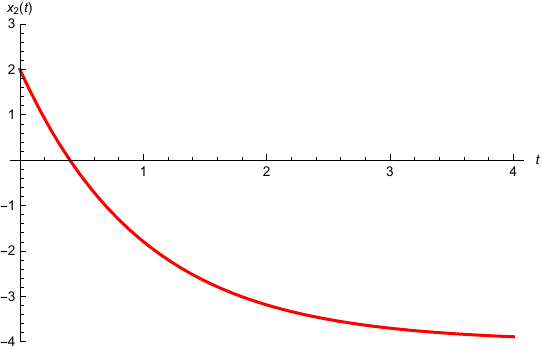}
		\caption{We plot the functions $x_{1}(t)$ and $x_{2}(t)$ of eqs. (\ref{solutions x(t) example 3}%
			).}		\label{Ex3fig1}
	\end{center}
\end{figure}

\begin{figure}
	\begin{center}
		\includegraphics[height=12cm]{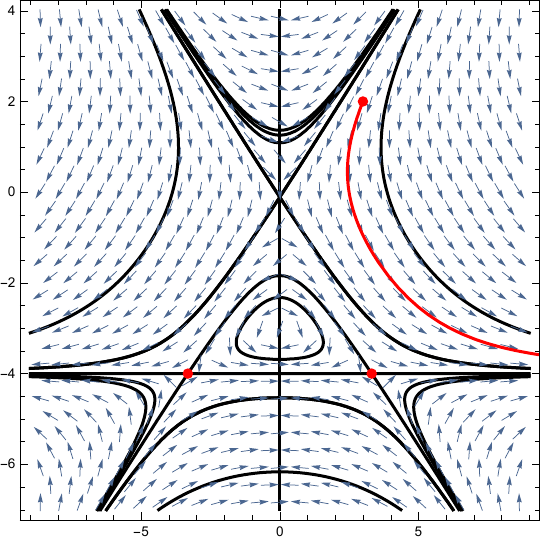}
		\caption{We plot in red the  \textit{trajectory} in the	$x_{1}-x_{2}$ plane corresponding to the solution (\ref{solutions x(t) example 3})
			together with several other\textit{ trajectories} in \textit{black}, with the
			$2$ \textit{ unstable equilibrium points} $\boldsymbol{\bar{x}^{(+)}}$  and $\boldsymbol{\bar{x}^{(-)}}$
			defined by eqs. (\ref{Example3 equilibrium points}) in red, the \textit{velocity}
			field $\boldsymbol{v}=\left(  \dot{x}_{1},\dot{x}_{2}\right)$, and the \textit{singularity line} $P_1(x_1,x_2)=x_1=0$. This solution (\ref{solutions x(t) example 3}) is well-defined for all values of the independent variable $t$, as it \textit{never} hits the \textit{singularity line}, and it can be extended indefinitely in the past and in the future. However, \textit{other trajectories} can  meet the \textit{singularity line} in a \textit{finite} time, as shown for instance by the lines drawn in \textit{black}.}		\label{Ex3fig2}
	\end{center}
\end{figure}

\bigskip

\subsection{Example 5}

In this fifth example we consider the case in which the parameters $\phi
^{(\pm)}$ defined by eq. (\ref{phi pm}) are \textit{real}, \textit{different} in modulus, and
have the \textit{same} signs. We set
\end{subequations}
\begin{align}
\eta_{11}  &  =1/2~,\eta_{22}=3/2~,\nonumber\label{eta parameters example 4}\\
\eta_{12}  &  =-1/4~,\eta_{21}=-1~,
\end{align}
so that%
\begin{align}
&  \phi^{(-)} =1-1/\sqrt{2}~,\quad\phi^{(+)} =1+1/\sqrt{2}~.
\end{align}
In this case the point $\boldsymbol{\bar{y}}=(u_{1},u_{2})= (5/2,1)$ in the $y_1-y_2$ Cartesian plane identifies an \textit{unstable} equilibrium solution of the linear system of ODEs (\ref{ydot}).

\begin{figure}
	\begin{center}
		\includegraphics[height=7cm]{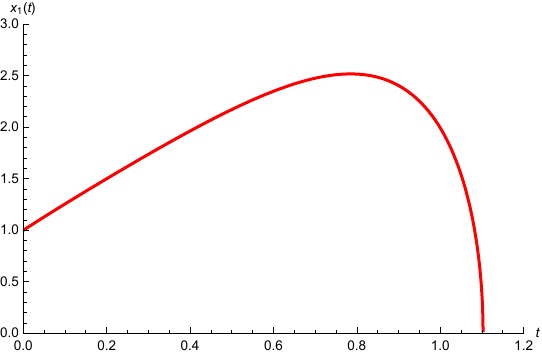}\\
		\includegraphics[height=7cm]{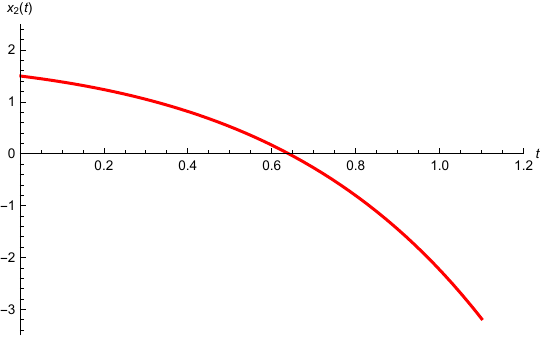}
		\caption{We plot the functions $x_{1}(t)$ and $x_{2}(t)$ of eqs. (\ref{solutions x(t) example 4}). The corresponding solution of the system of 2 ODEs (\ref{ODEs}) becomes \textit{undetermined} when $x_1(t)=0$.  }		\label{Ex4fig1}
	\end{center}
\end{figure}

\begin{figure}
	\begin{center}
		\includegraphics[height=11cm]{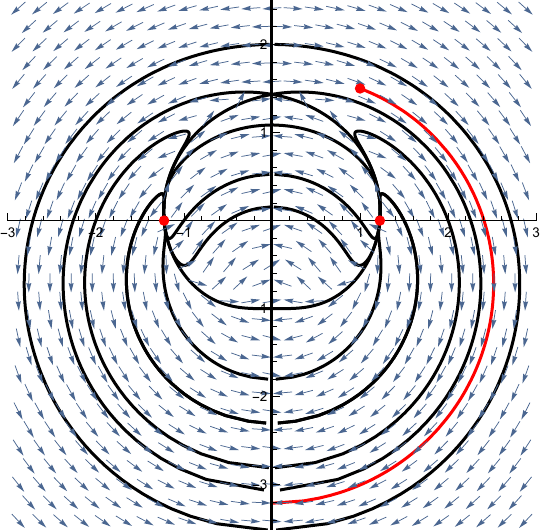}
		\caption{We plot in \textit{red}  the trajectory  in the $x_{1}-x_{2}$ plane corresponding to eqs. (\ref{solutions x(t) example 4}) , together with several	other solutions in \textit{black}, with the 2 unstable equilibrium points 
			$\boldsymbol{\bar{x}^{(+)}}$ and $\boldsymbol{\bar{x}^{(-)}}$ defined by eqs. (\ref{Example4 equilibrium points}),
			the velocity field $\boldsymbol{\bar{v}}= \left(  \dot x_{1}, \dot x_{2}\right)$, and the \textit{singularity line} $P_1(x_1,x_2)=x_1=0$. \textit{All} the trajectories meet the singularity line, indeed, in this case, \textit{all} the solutions of the system of ODEs (\ref{ODEs}) eventually cease to be solutions at the finite time $t_S$.
		}		\label{Ex4fig2}
	\end{center}
\end{figure}

We  take again $\lambda_{1}$, $\lambda_{2}$, and $\beta_{1}$ as in eqs.
(\ref{lambda h conditions}), so that $h_{0}=h_{2}=0$, and we choose the
following parameters as follows:%
\begin{subequations}
\label{parameters example 4}%
\begin{align}
\xi_{1}  &  =-1~,\xi_{2}=1~,\beta_{1}=0~,\beta_{2}=1~,\gamma=1~,\mu
=1~,\nonumber\\
\alpha_{0}  &  =1~,\alpha_{1}=0~,\alpha_{2}=1~, \lambda_{1} = 0~, \lambda
_{2}=1~, h_{1}=2~,
\end{align}
so that the eqs. (\ref{lambda h conditions}) give%
\begin{equation}
\lambda_{1}=0~,\lambda_{2}=1/2~,\beta_{1}=0~.
\end{equation}
We also have
\end{subequations}
\begin{subequations}
\label{Example 4 polynomials}%
\begin{align}
P_{3}^{(1)}(x_{1},x_{2})  &  =\left[ 8\left(  x_{1}\right)  ^{2}x_{2}+ 8\left(  x_{2}\right)  ^{3}+6\left(
x_{1}\right)  ^{2}+2\left(
x_{2}\right)  ^{2}-13x_{2}-9\right]  /4~,\\
P_{3}^{(2)}(x_{1},x_{2})  &  =\left[  -2\left(  x_{1}\right)  ^{2}-2\left(
x_{2}\right)  ^{2}+x_{2}+3\right]  x_{1}~,\\
P_{1}(x_{1},x_{2})  &  =2x_{1}~.
\end{align}
Again, $\dot{x}_{2}$ is \textit{never} singular, see eqs. (\ref{ODEs}).

Now we get $u_{1}=5/2$ and $u_{2}=1$, see eqs. (\ref{u1}) and (\ref{u2}), and the 2 points%

\end{subequations}
\begin{subequations}
\label{Example4 equilibrium points}%
\begin{equation}
\boldsymbol{\bar{x}^{(+)}}=\left(  X_{1}^{(+)}\left[  u_{1},u_{2}\right]  ,X_{2}^{(+)}\left[
u_{1},u_{2}\right]  \right)  = \left(  \sqrt{3/2} ,0\right)  \, ,
\end{equation}

\begin{equation}
\boldsymbol{\bar{x}^{(-)}}=\left(  X_{1}^{(-)}\left[
u_{1},u_{2}\right]  ,X_{2}^{(-)}\left[  u_{1},u_{2}\right]  \right)  = \left(
-\sqrt{3/2} ,0\right)
\end{equation}
of the $x_1-x_2$ Cartesian plane identify 2 \textit{unstable} equilibrium solutions of the system of ODEs (\ref{ODEs}), since the point
$\boldsymbol{\bar{y}}=(u_{1},u_{2})$ of the $y_1-y_2$ Cartesian plane  is itself an \textit{unstable} equilibrium solution of the system of ODEs (\ref{ydot}).

For the parameters in eqs. (\ref{parameters example 4}), we get

\end{subequations}
\begin{subequations}
\label{solution x example 4}%
\begin{align}
	\label{solution xx example 4}%
&  X_{1}^{(\pm)}\left[  y_{1}(t),y_{2}(t)\right]  =\pm\frac{1}{2}\left\{  10+
\exp\left(  t\right)  \left[  \sqrt{2}\left(  -6y_{1}\left(  0\right)
+y_{2}\left(  0\right)  +14\right)  \sinh\left(  t/\sqrt{2}\right)  \right.
\right. \nonumber\\
&  \left.  +2\left(  2y_{1}\left(  0\right)  +2y_{2}\left(  0\right)
-7\right)  \cosh\left(  t/\sqrt{2}\right)  \right] \nonumber\\
&  -\frac{1}{4\left(  198\sqrt{2}+283\right)  } \left[  -2\left(  11\sqrt
{2}+18\right)  \left(  2y_{1}\left(  0\right)  -y_{2}\left(  0\right)
-4\right)  \exp\left(  t\right)  \sinh\left(  t/\sqrt{2}\right)  \right.
\nonumber\\
&  \left.  \left.  +4\left(  9\sqrt{2}+11\right)  \left(  y_{2}\left(
0\right)  -1\right)  \exp\left(  t\right)  \cosh\left(  t/\sqrt{2}\right)
+36\sqrt{2}+44\right]  ^{2} \right\}  ^{1/2}\,,
\end{align}

\begin{align}
&  X_{2}^{(\pm)}\left[  y_{1}(t),y_{2}(t)\right]
=\left\{  \left[  9\sqrt{2}y_{2}\left(  0\right)  -2\left(  2\sqrt
{2}+7\right)  y_{1}\left(  0\right)  \right.  \right. \\
&  \left.  +11y_{2}\left(  0\right)  +\sqrt{2}+24\right]  \sinh\left(
\sqrt{2}t\right) \nonumber\\
&  \left.  +\left[  9\sqrt{2}y_{2}\left(  0\right)  -2\left(  2\sqrt
{2}+7\right)  y_{1}\left(  0\right)  +11y_{2}\left(  0\right)  +\sqrt
{2}+24\right]  \cosh\left(  \sqrt{2}t\right)  +\right. \nonumber\\
&  \left.  +2\left(  2\sqrt{2}+7\right)  y_{1}\left(  0\right)  +\left(
5\sqrt{2}-3\right)  y_{2}\left(  0\right)  -15\sqrt{2}-32\right\}  \frac
{\exp\left[  t\left(  1-\frac{1}{\sqrt{2}}\right)  \right]  }{14\sqrt{2}+8}
\,,
\end{align}
where the \textit{initial} values $y_{n}(0)$ are given by the expressions (\ref{y120}) in terms of the \textit{initial} values $x_{n}(0)$:

\end{subequations}
\begin{subequations}
\label{y120 Example 4}%
\begin{equation}
y_{1}\left(  0\right)  =\left[  x_{1}(0)\right]  ^{2}+\left[  x_{2}(0)\right]
^{2}+x_{2}(0)+1~,
\end{equation}%
\begin{equation}
y_{2}\left(  0\right)  =x_{2}(0)+1~.
\end{equation}
For initial values $x_{1}(0)=1$ and $x_{2}(0)=3/2$, we get $y_{1}(0)=23/4$
and $y_{2}(0)=5/2$. Since the initial value $x_{1}(0)$ is \textit{positive}, we have to
choose the \textit{positive} sign in (\ref{solution xx example 4}), so that the
corresponding solution is
\end{subequations}
\begin{subequations}\label{solutions x(t) example 4}
\begin{align}
x_{1}(t)  &  =\left\{  82 \left(  198 \sqrt{2}+283\right)  \exp\left(  2
t\right)  +\left(  3904 \sqrt{2}+5546\right)  \exp\left[  \left(  1+
{1}/{\sqrt{2}}\right)  t\right]  \right. \nonumber\\
&  \left.  +82 \left(  20 \sqrt{2}+29\right)  \exp\left[  \left(  1-
{1}/{\sqrt{2}}\right)  t\right]  -\left(  3192 \sqrt{2}+4817\right)
\exp\left[  \left(  2+\sqrt{2}\right)  t\right]  \right. \nonumber\\
&  \left.  -1681 \left(  12 \sqrt{2}+17\right)  \exp\left[  \left(  2-\sqrt
{2}\right)  t\right]  +4752 \sqrt{2}+6792 \right\}  ^{1/2} \nonumber\\
& \times \left({\sqrt{2}-1}/{(8
\sqrt{2}+28)}\right)\,,\\
x_{2}(t)  &  =\exp\left[  \left(  1-\frac{1}{\sqrt{2}}\right)  t\right]
\left[ \left(  \sqrt{2}-58\right)  e^{\sqrt{2} t}+41 \left(  \sqrt
{2}+2\right)\right]/\left[4 \left(  7 \sqrt{2}+4\right) \right] \,,
\end{align}

The functions $x_{1}(t)$ and $x_{2}(t)$ in eqs. (\ref{solutions x(t) example 4}) are plotted in Fig. \ref{Ex4fig1}. The corresponding trajectory in the
$x_{1}-x_{2}$ plane is plotted in \textit{red} in Fig. \ref{Ex4fig2}, together with
several other solutions in \textit{black}, with the 2 unstable equilibrium points
points $\boldsymbol{\bar{x}^{(+)}}$ and $\boldsymbol{\bar{x}^{(-)}}$ defined by eqs.
(\ref{Example4 equilibrium points}), the velocity field $\boldsymbol{v}=\left(
\dot{x}_{1},\dot{x}_{2}\right)  $, and the singularity line $P_{1}(x_{1}%
,x_{2})=0$. Also in this case, \textit{all} the trajectories hit the singularity line,
indeed, \textit{all} the solutions of the system of ODEs (\ref{ODEs}) eventually cease
to be solutions at the finite value $t=t_S$.
\end{subequations}
\bigskip

\section{A summary of our findings}

In this \textbf{Section 4} we provide a terse summary of our main findings.

Our main finding---which we believe to be new---is that the system of $2$
\textit{nonlinearly-coupled} ODEs (\ref{ODEs})---which features $23$ \textit{a
priori arbitrary} coefficients---is \textit{explicitly solvable} (as detailed
in \textbf{Section 2}), if these\textit{\ }$23$ \textit{a priori arbitrary}
coefficients satisfy $12$ \textit{constraints}, which we are able to identify
\textit{explicitly: }see \textbf{Subsection 2.4}. Moreover, the $2~$ \textit{additional constraints} in eq. (\ref{restriction eta}) (or rather\textit{\ "one and a half}", since
only one requires that an \textit{explicit equality} involving the
coefficients hold, while the other amounts only to an \textit{explicit
inequality}) imply that \textit{all} \textit{solutions} uniquely determined by the system (\ref{ODEs}) for \textit{all} values of the independent variable $t$ are periodic with the \textit{same} period (\textit{isochrony}!):
see \textbf{Subsection 2.3}. We also identified \textit{explicitly solvable} cases of the system of 2 first-order ODEs featuring in its right-hand sides 2 \text{cubic} polynomials:
\begin{equation}
	\dot{x}_{n}=P_{3}^{\left(  n\right)  }\left(  x_{1},x_{2}\right)
	 ~,~~~n=1,2~; \label{xxxdot}%
\end{equation}
we also reported other examples of solvable systems of two nonlinearly coupled first order ODEs displaying some peculiarities of their solutions, see  \textbf{Section 3}.

Finally let us mention that the techniques employed in this paper may of course be extended in several directions (although doing so shall not be a \textit{trivial} task): for instance, to systems of 2 first-order ODEs with \textit{more complicated} right-hand sides, to systems of \textit{more} than just 2 first-order ODEs (or, somewhat equivalently, to systems involving \textit{higher} than first-order derivatives); to system of \textit{PDEs}, involving more than a single independent variable; also to systems involving as dependent variables more complicated mathematical entities (for instance \textit{non commuting} objects such as \textit{matrices}); and even to \textit{non-autonomous} systems.

\bigskip

\section{Acknowledgements}

We like to thank Fernanda Lupinacci who, in these difficult times, with
extreme efficiency and kindness, managed all the arrangements necessary to
make possible the visits of FP (with her family) in Italy which were essential to facilitate our collaboration. FC---feeling that for obvious gerontological reasons he is approaching the end of his scientific carrier---would like to thank his 2 younger co-authors---this paper was indeed completed in difficult times for the world---as well as the many, mainly younger, other co-authors with whom he collaborated during his long life. And FP also would like to thank Payame Noor University for financial support, which made possible her participation to this research.

\end{document}